\documentclass[twocolumn]{bmcart}

\usepackage{comment}
\usepackage{amssymb}
\usepackage{amsmath}
\usepackage{xcolor}
\usepackage{graphicx}
\usepackage{float}
\usepackage{wrapfig}
\usepackage{subfigure}
\usepackage{cite}
\usepackage{hyperref}
\usepackage[utf8]{inputenc}
\usepackage{array}
\usepackage[switch]{lineno}

\newcolumntype{L}[1]{>{\raggedright\let\newline\\\arraybackslash\hspace{0pt}}m{#1}}
\newcolumntype{C}[1]{>{\centering\let\newline\\\arraybackslash\hspace{0pt}}m{#1}}
\newcolumntype{R}[1]{>{\raggedleft\let\newline\\\arraybackslash\hspace{0pt}}m{#1}}

\startlocaldefs
\endlocaldefs

\begin{document}

\begin{frontmatter}

\begin{fmbox}
\dochead{Research}


\title{An investigation into  inter- and intragenomic variations of graphic genomic signatures}

\author[
   addressref={aff1}, 
   email={rkaramic@uwo.ca}   
]{\inits{RK}\fnm{Rallis} \snm{Karamichalis}}
\author[
   addressref={aff1},
   corref={aff1},
   email={lila.kari@uwo.ca}
]{\inits{LK}\fnm{Lila} \snm{Kari}}
\author[
   addressref={aff2},
   email={s.konstantinidis@smu.ca}
]{\inits{SK}\fnm{Stavros} \snm{Konstantinidis}}
\author[
   addressref={aff1,aff2},
   email={steffen@csd.uwo.ca}
]{\inits{SK}\fnm{Steffen} \snm{Kopecki}}

\address[id=aff1]{
  \orgname{Department of Computer Science, University of Western Ontario}, 
  \city{London, ON},     
  \cny{Canada}          
}

\address[id=aff2]{
  \orgname{Department of Mathematics and Computing Science, Saint Mary's University},
  \city{Halifax, NS},
  \cny{Canada}
}


\begin{artnotes}
\end{artnotes}



\begin{abstractbox}

\begin{abstract} 
\parttitle{Background} 
Motivated by the general need to identify and classify species based on molecular evidence, genome comparisons have  been proposed that are based on
measuring Euclidean distances between Chaos Game Representation (CGR) patterns of genomic DNA sequences. 
 
\parttitle{Results} 
We  provide, on an extensive dataset and using several different distances,  confirmation of the hypothesis that CGR patterns are preserved along a genomic  DNA sequence, and  are different for DNA sequences originating from genomes of different species. This finding lends support to the theory that CGRs of genomic sequences can act as {\em graphic genomic signatures}.
In particular,  we compare the CGR patterns of over five hundred different 150,000~bp    genomic sequences originating  from the genomes of six  organisms, each belonging to one of the kingdoms of life:
 \textit{H. sapiens} (Animalia; chromosome  21), 
 \textit{S. cerevisiae} (Fungi; chromosome 4), 
 \textit{A. thaliana} (Plantae; chromosome 1), 
  \textit{P. falciparum} (Protista; chromosome 14),
 \textit{E. coli} (Bacteria - full genome),  and
 \textit{P. furiosus} (Archaea - full genome). 
 We also provide  preliminary evidence of this method's applicability to  closely related species  by comparing   \textit{H. sapiens} (chromosome 21) sequences   and over one hundred and fifty genomic sequences, also 150,000 bp long,  from \textit{P. troglodytes} (Animalia;  chromosome Y),  for a total length of more than  101 million basepairs analyzed. 
We compute pairwise distances between CGRs of these  genomic sequences  using six different distances, and  construct Molecular Distance Maps  that visualize all sequences as points in a two-dimensional or three-dimensional space,  to  simultaneously display their interrelationships. 
\parttitle{Conclusion} 
Our analysis  confirms  that CGR patterns of DNA sequences  from the same genome are in general quantitatively similar, while being different for DNA sequences from  genomes of different species.  Our analysis of the  performance of  the  assessed distances  uses three  different quality measures and suggests   that  several  distances outperform the Euclidean distance, which has so far been  almost exclusively used  for such studies. In particular we show that, for this dataset, DSSIM (Structural Dissimilarity Index) and the descriptor distance (introduced here) are best able to classify genomic sequences.

\end{abstract}



\end{abstractbox}
\end{fmbox}

\end{frontmatter}




\section*{Introduction}
Alongside DNA barcoding, \cite{barcoding} and Klee diagrams \cite{klee2010biodiversity},  Chaos Game Representation (CGR) patterns of genomic segments  have been proposed  as another method for the classification and  identification of  genomic sequences \cite{Jeffrey90,MapOfLife,Edwards2002,HIV_analysis,Deschavanne1999}. The concept of {\it genomic signature} was first introduced in \cite{karlin95}, as being any specific quantitative characteristic of a DNA genomic sequence that is  pervasive along the genome of the same organism, while being dissimilar for DNA sequences originating from different organisms. Initial studies \cite{Jeffrey90,Deschavanne2000}, suggested that short fragments of genomic sequences retain most of the characteristics of the species they come from,  thus implying that genomic signatures exist.  Moreover,  the Chaos Game Representation (CGR) of a DNA sequence, a graphic representation of its sequence composition, was proposed in \cite{Jeffrey90} as having both the pervasiveness and differentiability properties necessary for it to qualify as a genomic signature.  This hypothesis was  quantitatively tested  and largely confirmed in \cite{MapOfLife} for 3,176 mitochondrial DNA (mtDNA) sequences,  and Molecular Distance Maps  were proposed therein as  vizualizations of species relationships based on measuring the distances between the CGR-images of their mtDNA genomes. Note that CGR patterns of mtDNA sequences can be different from those of  DNA sequences  from the major genome of the same organism, and  that  large scale quantitative analyses of the hypothesis that CGR can play the role of  a genomic  signature for genomic  sequences have not, to our knowledge, been performed.  The objective of this study is to confirm that CGR  can play the role of genomic signature for genomic DNA sequences, as well as to assess various distances that can be used to compare CGRs of genomic sequences.

We  analyze  508 fragments,  150 kbp  (kilo base pairs) long, taken from  complete genomic  DNA sequences of six species, each representing a different kingdom: 
chromosome  21 of \textit{Homo~sapiens}, 
chromosome  4 of \textit{Saccharomyces~cerevisiae},
chromosome 1 of \textit{Arabidopsis~thaliana}, 
chromosome 14 of \textit{Plasmodium~falciparum},
the genome of  \textit{Escherichia~coli},  
and the genome of
 \textit{Pyrococcus  furiosus}, for a total length of  76,200,000 bp analyzed.  We analyze the intergenomic and intragenomic variation of CGR genomic signatures of these sequences by using six different distances for image comparison: Structural Dissimilarity Index (DSSIM) \cite{ssim}, Euclidean distance, Pearson correlation distance \cite{pearson}, Manhattan distance \cite{manhattan},  approximated information distance \cite{Li2004}, and a distance we propose here, called {\it descriptor  distance}.  We visualize the results by computing the Molecular Distance Maps of all  DNA sequences in the database, for each of the six distances. The resulting Molecular Distance Maps show  a good clustering of the DNA sequences, with those originating from the same genome being largely grouped together, and separated from sequences belonging to genomes of different organisms. We observe that, in some of the cases where  the clustering was suboptimal, the computation of three-dimensional Molecular Distance Maps resolves what appeared to be cluster overlaps in the two-dimensional Molecular Distance Maps. Lastly, using the ``ground-truth'' that sequences from the same genomes should have similar structural characteristics and thus be grouped together, while those from genomes of different organisms  should be separated, we  assess the six distances by combining three different quality measures: correlation to an idealized cluster distance,   silhouette accuracy, and histogram overlap. We conclude that DSSIM and the descriptor distance perform best according to these measures.  We also provide  preliminary evidence of this method's applicability to classifying  genomic DNA sequences of closely related species by comparing  the {\it H. sapiens} (chromosome 21)  sequences  with 168   genomic DNA sequences, 150 kbp long, from \textit{Pan  troglodytes} (chimp, chromosome Y),  for an  additional length of 25,200,000  bp analyzed. 
 Further research  may lead to  improvements  of these distances for   optimal genomic DNA sequence identification and classification results.

Note that other alignment-free methods have been  used for phylogenetic analysis of DNA sequences.
The initial reports  on CGRs of genomic sequences \cite{Jeffrey90, Jeffrey92} contained mostly qualitative assessments of CGR patterns of whole genes. In \cite{Deschavanne1999},  several datasets of up to 36  genomic DNA sequences were analyzed,  and in \cite{Deschavanne2000} some various-length  sequences were analyzed based on computing  Euclidean distances between frequencies of their $k$-mers,  for $k=1,..., 8$. Subsequently,  \cite{Edwards2002}   computed the Euclidean distance between frequencies of $k$-mers ($k \leq 5$) for the analysis of  125 GenBank DNA sequences from 20 bird species and the American alligator. In \cite{Pride2003},  27 microbial genomes  were analyzed  to find  implications of  4-mer frequencies  ($k=4$) on their evolutionary relationships.  In \cite{Li2004}, 20  mammalian  complete mtDNA  sequences were analyzed using  the ``similarity metric'', for $k=7$.  Another study, \cite{Deschavanne2010},  analyzed  459 bacteriophage genomes and compared them with their host genomes to infer  host-phage  relationships,  by computing Euclidean distances  between frequencies of $k$-mers  for $k=4$.  In \cite{Pandit2010}, 75  complete HIV genome sequences were compared using the Euclidean distance between frequencies of 6-mers ($k=6$), in order to group them in subtypes. In \cite{MapOfLife} a  dataset of 3,176  complete mtDNA sequences was analyzed, and several Molecular Distance Maps were obtained using DSSIM  and a value of  $k=9$.

The main contributions of this paper are:
\begin{itemize}
\item  We tested and confirmed for an extensive dataset, of a total length of  101,400,000 bp, the hypothesis that CGR images of {\it genomic} DNA sequences can play the role of a {\it (graphic) genomic signature}, meaning that they have a desirable genome- and species- specificity.  The  dataset comprised 150 kbp long sequences taken from genomes of organisms from each of the six kingdoms of life,  augmented by   a set of same-length genomic sequences from \textit{P. troglogytes} as a test-case of this method's applicability to closely related species.

\item We assessed  the performance of six different distances in this context, and this analysis included both same-genome and different-genome DNA fragment pairs. For several of these distances,  the intragenomic values were overall smaller than intergenomic values,  suggesting that  this method  could    separate DNA genomic fragments  belonging to different  genomes, based on  their CGRs. 

 \item  We showed  that several distances outperform the Euclidean distance, which has so far been almost exclusively used for such  studies. In particular,  we determined that  the DSSIM distance and descriptor distance (introduced here), both of whom essentially compare the $k$-mer composition of DNA sequences (herein $k=9$), were best able to differentiate sequences originating from different genomes in this dataset.  
 
 \item This study represents, to the best of our knowledge,  the largest combined dataset size and value of $k$ for this type of analysis. 

 \item Based on preliminary data, we suggest the use of three-dimensional Molecular Distance Maps for improved visualization of the simultaneous interrelationships among  similar or very distant DNA sequences.

\end{itemize}

\section*{Methods}

In this section we first describe the dataset used for our analysis, then present an overview of  the three  main steps of the method, and conclude with a description of the six distances that we considered.

\subsection*{Dataset}
\label{sec:dataset}

The dataset we  used includes complete genomic sequences from six organisms, each  representing  one of the six kingdoms of life, see Table~\ref{table1}. For additional information about the dataset see Appendix A.

\begin{table}[ht]
\centering
\begin{tabular}{| c | l | l | }
\hline & Organism  & NCBI Acc. Nr. \\ \hline
1 & \textit{H.~sapiens}, chrom. 21 (Animalia)& 	NC\_000021.8  \\
2 & \textit{E.~coli} (Bacteria)& NC\_000913.3 \\
3 & \textit{S.~cerevisiae}, chrom. 4 (Fungi)& NC\_001136.10 \\
4 & \textit{A.~thaliana}, chrom. 1 (Plantae)& NC\_003070.9  \\
5 & \textit{P.~falciparum}, chrom. 14 (Protista)& NC\_004317.2  \\
6 & \textit{P.~furiosus} (Archaea)& NC\_018092.1  \\ \hline
\end{tabular}
\vspace{3mm}
\caption{NCBI accession numbers of the dataset of the complete genomic DNA sequences considered, in increasing order of their NCBI accession number.}
\label{table1}
\end{table}

\begin{table}[ht]
\centering
\begin{tabular}{ | l | r | r | r |}
\hline
 Organism  & Length(bp) & \# Letters ``N''  & \# Fragments \\ \hline
\textit{H.~sapiens}& 48,129,895 & 13,023,253 & 234 \\
\textit{E.~coli} & 4,641,652 & 0 & 30 \\
\textit{S.~cerevisiae}& 1,531,933 & 0 & 10 \\
\textit{A.~thaliana}& 30,427,671 & 164,359 & 201 \\
\textit{P.~falciparum}& 3,291,871 & 37 & 21 \\
\textit{P.~furiosus}& 1,909,827 & 10 & 12 \\ \hline
\end{tabular}
\vspace{3mm}
\caption{Organism  considered, total length of genomic sequence, number of ignored letters ``N'', and number of DNA fragments (sequences) obtained by splitting each complete genomic DNA sequence into consecutive, non-overlapping, equal length (150 kbp) contiguous fragments.}
\label{table2}
\end{table}

In order to have relatively comparable number of DNA sequences for each organism, we chose the longest chromosomes for all organisms except {\it H.~sapiens}, for which the shortest chromosome was chosen.

The DNA sequences in the NCBI database are represented as strings of letters ``A'', ``C'', ``G'',  ``T'',  and ``N'' which represent the four nucleobases Adenine, Cytosine, Guanine,  Thymine,  and ``unidentified Nucleotide'', respectively.
For our analysis we  ignored all letters ``N''.  In \textit{S.~cerevisiae} and \textit{E.~coli} there were no ignored letters, and in \textit{P.~falciparum} and \textit{P.~furiosus} the number of ignored letters is of the order of $0.001\%$ of the length of the sequence. In \textit{H.~sapiens} this number is $27\%$, and in \textit{A.~thaliana} is $0.54\%$. In \textit{H.~sapiens}, in particular, $96.4\%$ of these ignored letters exist in centromeric and telomeric regions of the chromosome.

The resulting genomic DNA sequences were divided into  successive, non-overlapping, contiguous fragments, each 150 kbp long. When the last sequence   was shorter than 150 kbp, it was not included in the analysis.  This resulted in
 $234$ fragments for  \textit{H. sapiens}, 
 $30$ fragments  for  \textit{E. coli}, 
 $10$  fragments for \textit{S. cerevisiae}, 
 $201$ fragments  for  \textit{A. thaliana},
 $21$  fragments for  \textit{P. falciparum},
and $12$ fragments for  \textit{P. furiosus}, for a total of 508 DNA fragments,
see Table~\ref{table2}.

\subsection*{Overview}
\label{sec:overview}

The method we used to analyze and classify the 508 sequences of the dataset   has three steps:  {\it (i)} generate graphical representations (images)  of each DNA sequence using Chaos Game Representation (CGR), {\it (ii)} compute all pairwise distances between  these images, and {\it (iii)}  visualize the interrelationships implied by these distances  as  two- or three-dimensional maps,  using Multi-Dimensional Scaling (MDS).

\begin{figure}[h]
\centering
\subfigure[\textit{H.~sapiens}]{
    \includegraphics[width=0.27\linewidth]{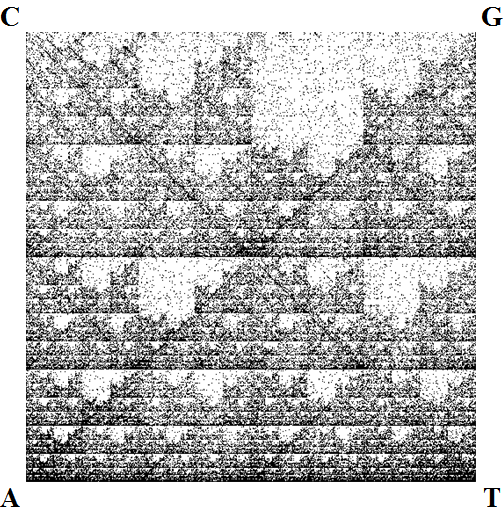}
    \label{fig:subfig1}
}
\subfigure[\textit{E.~coli}]{
    \includegraphics[width=0.27\linewidth]{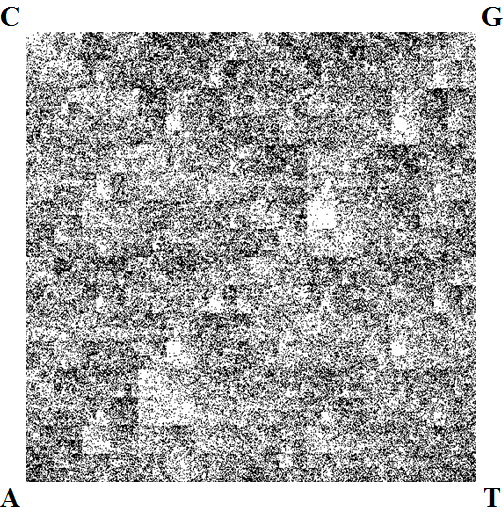}
    \label{fig:subfig5}
}
\subfigure[\textit{S.~cerevisiae}]{
    \includegraphics[width=0.27\linewidth]{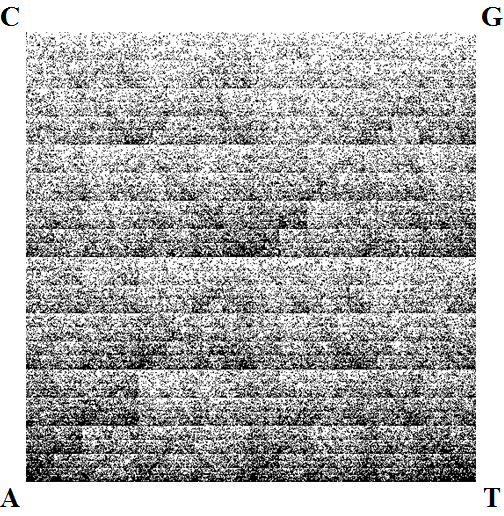}
   \label{fig:subfig2}
}\\
\subfigure[\textit{A.~thaliana}]{
    \includegraphics[width=0.27\linewidth]{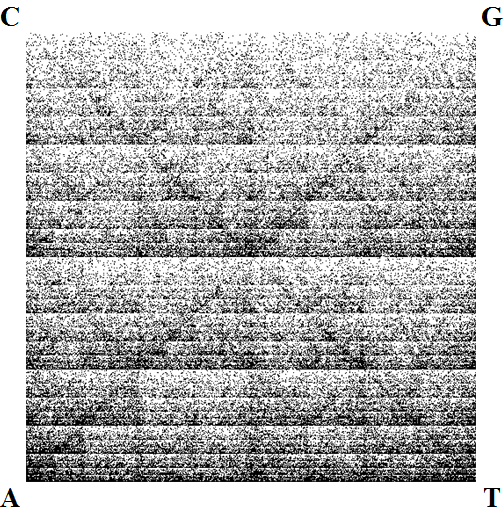}
    \label{fig:subfig3}
}
\subfigure[\textit{P.~falciparum}]{
    \includegraphics[width=0.27\linewidth]{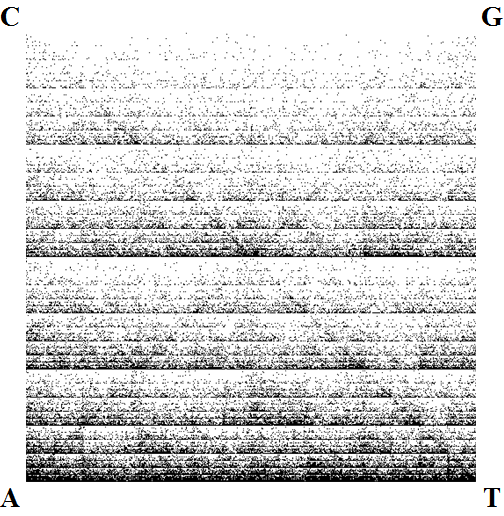}
    \label{fig:subfig4}
}
\subfigure[\textit{P.~furiosus}]{
    \includegraphics[width=0.27\linewidth]{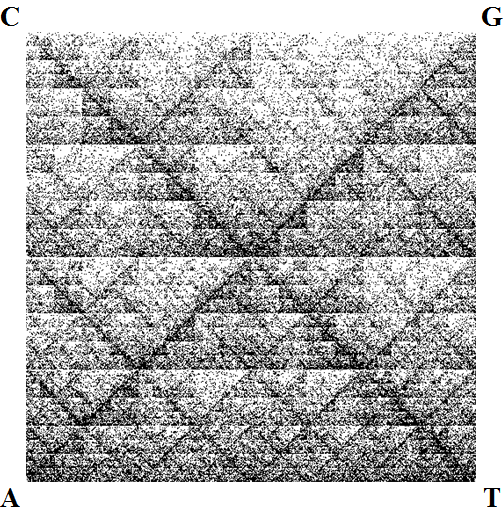}
    \label{fig:subfig6}
}
\caption{$2^9 \times 2^9$ CGR images of   150 kbp genomic DNA  sequences.
 of \textit{H.~sapiens}, \textit{E.~coli}, \textit{S.~cerevisiae},  \textit{A.~thaliana}, \textit{P.~falciparum},  and \textit{P.~furiosus}.
}
\label{fig:all_fcgrs}
\end{figure}

CGR is  a method  introduced by Jeffrey \cite{Jeffrey90} in 1990 to visualize the structure of a DNA sequence.
A CGR  associates an image to each DNA sequence as follows. Starting from a unit square with corners labelled {\it A, C, G,} and {\it T}, and the center of the square as the starting point, the image is obtained by  successively plotting each nucleotide as the middle point between the current point and the  corner labelled by the nucleotide to be plotted. If the generated square image has a size of $2^k\times 2^k$ pixels, then every pixel represents a distinct $k$-mer: A pixel is black if the $k$-mer it represents occurs in the DNA sequence, otherwise it is white.  CGR images of genetic DNA sequences originating from various species show  patterns such as squares, parallel lines, rectangles, triangles, and also complex fractal patterns, Figure~\ref{fig:all_fcgrs}.

For step {\it (i)},  a slight modification of the original CGR was used,  introduced by Deschavanne \cite{Deschavanne1999}: a $k$-th order FCGR  (frequency CGR) is a $2^k \times 2^k$  matrix that can be constructed  by dividing the CGR plot into a $2^k \times 2^k$ grid, and defining the element $a_{ij}$ as the number of points that are situated in  the corresponding grid square.  A first and second order FCGR are shown below, where $N_{w}$ is the number of occurrences of the oligonucleotide $w$ in the sequence $s$.

$$
FCGR_{1}(s) =
\left(
\begin{array}{cc}
N_{C} & N_{G} \\ 
N_{A} & N_{T} \\
\end{array}
\right), $$
$$  
FCGR_{2}(s) =
\left(
\begin{array}{cccc}
N_{CC} & N_{GC} & N_{CG} & N_{GG} \\
N_{AC} & N_{TC} & N_{AG} & N_{TG} \\
N_{CA} & N_{GA} & N_{CT} & N_{GT} \\
N_{AA} & N_{TA} & N_{AT} & N_{TT} \\
\end{array}
\right).
$$

The $(k+1)$-th order $FCGR_{k+1}(s)$ can be obtained by replacing each element $N_{X}$ in
$FCGR_{k}(s)$ with four elements
$$
\left(
\begin{array}{cc}
N_{CX} & N_{GX} \\
N_{AX} & N_{TX} \\
\end{array}
\right)
$$

\noindent
where $X$ is a sequence of length $k$ over the alphabet $\{A, C, G, T\}$.

For step {\it (ii)}, after computing the FCGR matrices for each of the 150 kbp sequences in our dataset, the goal was to measure  ``distances'' between every two CGR images. There are many distances that can be defined and used  for this purpose, \cite{deza2009}.  One of the goals of this study was to identify what distance  is better able to differentiate the structural differences of various genomic DNA sequences and  classify them  based on the species they belong to.
 In this paper we use six different distances:  Structural Dissimilarity Index (DSSIM), 
 descriptor distance (defined here), Euclidean distance, Manhattan distance, Pearson correlation distance, and approximated information distance. 

For step {\it (iii)}, after computing all possible pairwise distances we obtained six different distance  matrices. To visualize the inter-relationships between sequences implied by  each of the distance matrices, and to  thus  visually assess  each of the distances,  we used  Multi-Dimensional Scaling (MDS). MDS is  an information visualization  technique introduced by Kruskal in \cite{kruskal}. Given as input a distance matrix that contains the pairwise distances among a set of items\footnote{In this paper the items are the  150 kpb DNA sequences analyzed.},  the output of MDS is a  spatial representation of the items on a common Euclidean space wherein 
each item is represented as a point and the spatial distance between any two points corresponds to the distance between the items in the distance matrix:  Objects  with a small pairwise distance will result in  points that are close to each other,  while objects with a large pairwise distance  will become points that are far apart.  For example, in \cite{MapOfLife} MDS was used in conjunction with DSSIM and CGR to produce Molecular Distance Maps  that visually display the  simultaneous interrelationships among a set of full mitochondrial DNA sequences.
 
 The ideal Molecular Distance Map is a placement of $n$ items as points in an $(n-1)$-dimensional space. The two-dimensional Molecular Distance Map is simply an approximation, a flattening of this highly-dimensional space onto the plane, which may sometimes result in  erroneous positioning of some points. Increasing the dimensionality of the Molecular Distance Map often results in a more accurate representation  of the real interrelationships between sequences, as embodied in the original distance matrix.

\subsection*{Distances}
\label{sec:distances}

In this section we describe and formally define each of the six distances used in our analysis: DSSIM,  descriptor distance (introduced here), Euclidean, Manhattan, Pearson, and approximated information distance.

Structural Similarity Index, SSIM, was introduced in \cite{ssim} for the purpose of  assessing the degree of  similarity between two images. Given two images $X, Y$ as  $n \times n$  matrices having as elements integers ranging in the interval $[0, L]$, SSIM computes three factors (luminance, contrast and structure) and combines them to obtain a similarity value. However, instead of computing a global similarity between the two images,  each image is divided into  $11 \times 11$ sliding square windows $X^{ij} (Y^{ij}  \text{respectively})  \ \text{with} \ i,j=1, \cdots, n-10$ which move pixel by pixel to eventually cover  the entire image, and the SSIM similarity of any given pair of images  is computed by  comparing their corresponding windows. In addition, an $11 \times 11$ circular symmetric Gaussian weighting function $W \in \mathbb{R}^{11 \times 11}$ with a fixed standard deviation of $1.5$, normalized to unit sum ($\sum_{p=1}^{11}\sum_{q=1}^{11} W_{pq}=1$), is used. Then, the mean $\mu_{x,i,j}$ ($\mu_{y,i,j}$ for $Y$), variance $\sigma_{x,i,j}$ ($\sigma_{y,i,j}$ for $Y$) and correlation $\sigma_{xy,i,j}$ 
are computed, as follows:

$$
\mu_{x,i,j}=\sum_{p=1}^{11} \sum_{q=1}^{11}W_{pq}X^{ij}_{pq}
$$
$$
\sigma_{x,i,j}=\sqrt{\sum_{p=1}^{11} \sum_{q=1}^{11}W_{pq}(X^{ij}_{pq}-\mu_{x,i,j})^{2}}
$$
$$
\sigma_{xy,i,j}=\sum_{p=1}^{11} \sum_{q=1}^{11}
W_{pq}(X^{ij}_{pq}-\mu_{x,i,j})(Y^{ij}_{pq}-\mu_{y,i,j})
$$

\noindent
where  $A_{pq}$ denotes the $(p,q)$ element of the matrix $A$. Based on these values, the luminance $l(X^{ij},Y^{ij})$, contrast $c(X^{ij},Y^{ij})$ and structure $s(X^{ij},Y^{ij})$ are computed as

$$
l(X^{ij},Y^{ij})=\frac{2\mu_{x,i,j}\mu_{y,i,j}+C_{1}}{\mu_{x,i,j}^{2}+\mu_{y,i,j}^2+C_{1}}
$$
$$
c(X^{ij},Y^{ij})=\frac{2\sigma_{x,i,j}\sigma_{y,i,j}+C_{2}}{\sigma_{x,i,j}^{2}+\sigma_{y,i,j}^2+C_{2}}
$$
$$ 
s(X^{ij},Y^{ij})=\frac{\sigma_{xy,i,j}+C_{3}}{\sigma_{x,i,j}\sigma_{y,i,j}+C_{3}}
$$

where $C_{1}=(0.01)^{2}$, $C_{2}=(0.03)^{2}$, $C_{3}=\frac{C_{2}}{2}$. Then, these three factors are combined to get

$$
SSIM(X^{ij},Y^{ij})=l(X^{ij},Y^{ij}) c(X^{ij},Y^{ij})s(X^{ij},Y^{ij})
$$

and finally,  the  SSIM  index  used to evaluate the overall image similarity is computed as

$$
SSIM(X,Y)=\frac{1}{(n-10)^2}\sum_{i=1}^{n-10}\sum_{j=1}^{n-10} SSIM(X^{ij},Y^{ij}).
$$ 

In theory, the values for SSIM  range  in the interval $[-1,1]$ with the similarity being $1$ between two identical images, $0$, for example, between a black image and a white image, and $-1$ if the two images are negatively correlated; that is, SSIM$(X,Y)=-1$ if and only if $X$ and $Y$ have the same luminance $\mu$ and every pixel $x_i$ of image $X$ has the inverted value of the corresponding pixel $y_i = 2\mu - x_i$ in $Y$. 

To compute the distance rather than the similarity between two images, we calculate
 DSSIM $(X,Y) = 1-$ SSIM$(X,Y)$. Consequently, the range of  DSSIM  is the interval $[0,2]$:  two 
 identical images  will result  in a DSSIM distance of $0$, while two  images that are the negatives of each other  would result in a DSSIM distance of $2$.
 
The {\it descriptor distance} between  two FCGRs $X,Y \in \mathbb{N}^{2^k \times 2^k}$  aims to compare a combination of several  different``descriptors'', that is, a combination of several different aspects, of the two given FCGRs.

A {\it descriptor} is  a vector  characterized by  parameters $m$ and  $r$, as well as $r$ intervals, where $m$ is the size of the non-overlapping windows in which the FCGR is divided  (scale of the comparison), and the $r$  intervals represent the  ``granularity'' of the analysis, in that they define the  intervals  of  numbers of $k$-mer occurrences that are considered significant.

For a given  $m \leq k$ and $r$, and intervals  $[a_{0}, a_1), [a_{1}, a_{2}),$ $\cdots ,$ $ [a_{r-1}, a_r)$ such that $\bigcup_{i=0}^{r-1} [a_{i},a_{i+1})=[0,\infty)$ and $[a_{i},a_{i+1})\cap [a_{j},a_{j+1})= \emptyset$ $\forall i,j$ with $i \neq j$,  a decriptor is constructed as follows. 

Starting from the top-left corner,  we divide each of the two FCGR matrices $X$ and $Y$ into non-overlapping submatrices\footnote{In general, these windows (submatrices) can be overlapping, but in this paper we made the choice of using non-overlapping windows.}   of size $2^m \times 2^m$. This procedure results in $ 4^{k-m}$ submatrices $X_{ij}$ and $Y_{ij}$ with $i,j=1, \cdots ,2^{k-m}$, which will be pairwise compared.

The choice of the $r$ intervals, called ``bins'',  points to the fact that, rather than considering the finest granularity, we are interested in a coarser comparison. This means that,  instead of a computationally expensive pairwise comparison of all possible numbers  of occurrences of $k$-mers,  we are interested only in certain ``bins'' of such numbers. For example, in our case, we use $r=5$ and consider  only  5 different bins, that is only  $k$-mers with number of occurences: 0 (not occurring), 1 (one occurrence),  2 (two occurrences),  between 2 and 5, between 5 and 20, and greater than 20 (most frequent). Formally, we use $r=5$ and $[0,\infty)=[0,1)\cup[1,2)\cup[2,5)\cup[5,20)\cup	[20,\infty)$ as the  5 bins.

Afterwards, we compute for every $X_{ij}$ a vector vec$X_{ij} = \frac{1}{(2^m \times 2^m)}  (b_1, b_2, \cdots , b_r)$ where $b_i=| \{x \in X_{ij}: a_{i-1} \leq x <a_{i}\}|$.  In our case, for each $X_{ij}$, we  compute a five-tuple wherein, for example, the 4th element   represents the number of $9$-mers whose number of occurrences is in the 4th  bin, that is, at least   5 but less than 20.  The division to $2^m \times 2^m$ is to obtain a probability distribution for each submatrix.
The same procedure is performed for  $Y_{ij}$, resulting in the vector vec$Y_{ij}$.

We  further append all vectors vec$X_{ij}$ and form a new  vector vec$X^{m, r}$ and, using the same order of appending, we append all vectors vec$Y_{ij}$ forming a new vector vec$Y^{m, r}$.  These two vectors are the ``descriptors'' of the FCGR matrices $X$ and $Y$ for the parameters $m$, $r$ and the $r$ chosen bins.

As  a last step,  we combine descriptors vec$X^{m, r}$ (respectively  vec$Y^{m, r}$) for several values of $m$ and $r$  by appending  them one after another, in the same order, to obtain the vector  vec$X$ (respectively vec$Y$).

The {\it  descriptor distance} between the two FCGRs $X$ and $Y$  is now defined as the Euclidean distance between the vectors vec$X$ and vec$Y$

$$
d_D(X,Y)= d_E(\text{vec}X, \text{vec}Y).
$$

In our case   we  computed descriptors for $m=4,5,6$ therefore forming vectors vec$X$ and vec$Y$ of length  
$5\big((\frac{512}{64})^2+(\frac{512}{32})^2+(\frac{512}{16})^2\big)=6720$.   In general,  for a given $r$, the length of the  vectors compared is 
$ r ((2^{k-m_1})^2 + (2^{k-m_2})^2 + ... + (2^{k-m_p})^2)$, where $m_1, m_2, \dots, m_p$ are the values used for $m$. The choice of $m$ for this study was made  to balance the computational cost of calculating the vector of descriptors  with the ability  to compare the two matrices at various scales:  large  ($m=6$, that is, compare windows  of size  $64 \times 64$), medium ($m=5$, windows of size $32 \times 32$)) and   small  ($m=4$,  windows of size $16 \times 16$). The parameter $r=5$ and the 5 bins were kept constant throughout our calculations but, in general, these parameters  can also be varied,  and the resulting vectors for each value added to the vector of descriptors, resulting in a larger vector.

In principle, the descriptor distance between two  FCGRs  effectively compares   the distribution of frequencies of $k$-mers  between the corresponding submatrices  $X_{ij}$ and $Y_{ij}$, and does that for several values of $m$, that is, at several different scales. (Note that, in each window $X_{ij}$, all $k$-mers have the same suffix of length $k-m$.)

We  now illustrate the {\it descriptor distance}  by an example wherein  $k=3$, $m=2$,  $r=3$, and the 3 bins are $[0,15) \cup [15,30) \cup [30, \infty)$. Since $k=3$, the FCGR table will contain the
number of occurrences of all 3-mers in a DNA sequence, as follows:

\begin{table}[H]
\centering
\begin{tabular}{|@{\,}c@{\,}|@{\,}c@{\,}|@{\,}c@{\,}|@{\,}c@{\,}||||@{\,}c@{\,}|@{\,}c@{\,}|@{\,}c@{\,}|@{\,}c@{\,}|} \hline
 CCC & GCC & CGC & GGC & CCG & GCG & CGG & GGG \\   \hline  
 ACC & TCC & AGC & TGC & ACG & TCG & AGG & TGG \\   \hline 
 CAC & GAC & CTC & GTC & CAG & GAG & CTG & GTG \\   \hline  
 AAC & TAC & ATC & TTC & AAG & TAG & ATG & TTG \\   \hline \hline \hline \hline
 CCA & GCA & CGA & GGA & CCT & GCT & CGT & GGT \\   \hline   
 ACA & TCA & AGA & TGA & ACT & TCT & AGT & TGT \\   \hline
 CAA & GAA & CTA & GTA & CAT & GAT & CTT & GTT \\   \hline
 AAA & TAA & ATA & TTA & AAT & TAT & ATT & TTT \\   \hline 
\end{tabular}
\end{table}

Take the  two FCGRs $X, Y \in \mathbb{N}^{8 \times 8}$, ($k=3$, thus $2^3 \times 2^3$)  corresponding to two genomic 150 kbp sequences of our dataset (one human and one bacterial), respectively. In order to use small numbers throughout the example, we divide all  elements of the obtained matrices by $100$ and take the integer part of each element, obtaining:
 
$$
X=
\left(
\begin{array}{cccccccc}
 42 & 33 & 9 & 33 & 14 & 10 & 15 & 45 \\
 22 & 30 & 26 & 25 & 9 & 5 & 37 & 37 \\
 32 & 21 & 33 & 19 & 44 & 35 & 41 & 35 \\
 17 & 9 & 13 & 21 & 23 & 10 & 22 & 18 \\
 37 & 26 & 6 & 32 & 34 & 24 & 9 & 23 \\
 29 & 24 & 31 & 27 & 19 & 27 & 18 & 28 \\
 21 & 23 & 10 & 9 & 19 & 17 & 21 & 15 \\
 35 & 15 & 14 & 14 & 19 & 12 & 17 & 30 \\
\end{array}
\right),
$$
$$ 
Y=
\left(
\begin{array}{cccccccc}
 18 & 34 & 40 & 27 & 30 & 36 & 27 & 12 \\
 27 & 18 & 27 & 32 & 24 & 23 & 15 & 23 \\
 24 & 17 & 13 & 17 & 36 & 12 & 32 & 18 \\
 27 & 17 & 28 & 26 & 18 & 8 & 22 & 25 \\
 32 & 32 & 23 & 16 & 16 & 25 & 23 & 22 \\
 20 & 29 & 18 & 25 & 16 & 16 & 15 & 17 \\
 25 & 25 & 7 & 16 & 26 & 27 & 20 & 25 \\
 32 & 21 & 20 & 21 & 25 & 18 & 27 & 34 \\
\end{array}
\right).
$$

Thus, in the human DNA sequence, the triplet CCC appears about 4200 times, the triplet GCC appears about 3300 times, the triplet CGC appears about 900 times, etc.

Since $m=2$, we divide each of the matrices $X$ and $Y$ into non-overlapping submatrices of size $4 \times 4$ ($2^2 \times 2^2$). For $X$ we thus obtain $X_{11}, X_{12}, X_{21},X_{22}$

$$
\left(
\begin{array}{cccc}
 42 & 33 & 9 & 33 \\
 22 & 30 & 26 & 25 \\
 32 & 21 & 33 & 19 \\
 17 & 9 & 13 & 21 \\
\end{array}
\right), 
\left(
\begin{array}{cccc}
 14 & 10 & 15 & 45 \\
 9 & 5 & 37 & 37 \\
 44 & 35 & 41 & 35 \\
 23 & 10 & 22 & 18 \\
\end{array}
\right),
$$
$$
\left(
\begin{array}{cccc}
 37 & 26 & 6 & 32 \\
 29 & 24 & 31 & 27 \\
 21 & 23 & 10 & 9 \\
 35 & 15 & 14 & 14 \\
\end{array}
\right), 
\left(
\begin{array}{cccc}
 34 & 24 & 9 & 23 \\
 19 & 27 & 18 & 28 \\
 19 & 17 & 21 & 15 \\
 19 & 12 & 17 & 30 \\
\end{array}
\right).
$$
and similarly for $Y$.

Since the  $r=3$ bins  are $[0,15) \cup [15,30) \cup [30, \infty)$,  we will count, for each submatrix,  the number of $3$-mers for which the number of occurrences is less than $15$, between $15$ and $30$, and greater than  or equal to $30$. Thus we obtain vec$X_{11} = \frac{1}{16}(3, 7, 6)$ which has as elements the number of elements of $X_{11}$ which belong in each of the intervals selected, divided by the total number of elements of $X_{11}$. We proceed similarly for vec$X_{12} = \frac{1}{16} (5, 4,7)$, vec$X_{21} =  \frac{1}{16} (5, 7, 4)$, vec$X_{22} = \frac{1}{16}(2, 12, 2)$ and we form vec$X$ by appending these vectors one after the other, that is 

$$\textstyle
	\text{vec}X = \frac{1}{16}\left(3, 7, 6, 5 , 4, 7, 5, 7, 4, 2, 12, 2\right).
$$
We apply exactly the same procedure for the matrix $Y$ and we get 

$$\textstyle
	\text{vec}Y = \frac{1}{16} \left(1, 12, 3, 3, 9, 4, 1, 12,3 ,0, 15, 1\right).
$$
The descriptor distance between these two FCGRs is computed as the Euclidean distance between vec$X$ and vec$Y$, in this case $d_D(X,Y) \approx 0.718$. Note that, since we started by dividing the number of 3-mer occurrences by 100, as well  as  because of the bin selection, this is a fictitious example. The real value of the descriptor distance between the  mentioned human and bacterial sequences is  8.66, and the range of the descriptor distance for this dataset of DNA sequences is  [0, 13.17]. In general, the descriptor distance  has a variable range, that depends on the choices of parameters used.

To compute the Euclidean, Manhattan and Pearson distances, we first convert the matrices $X, Y \in \mathbb{N}^{n \times n} $ into $1 \times n^{2}$ vectors. For two vectors $x, y \in \mathbb{R}^n$, their Euclidean distance $d_E(x, y)$  and  their Manhattan distance $d_M(x, y)$ are computed as

$$  
d_E(x,y)=\sqrt{\sum _{i=1}^n (x_i -y_i)^2},
$$
$$
d_M(x,y)=\sum _{i=1}^n |x_i -y_i |,
$$
while their Pearson distance $d_P(x, y)$ is defined as 
$$
d_P(x,y)=1-\frac{\sigma_{xy}}{\sigma_x \sigma_y},
$$
where
$$
\mu_x=\frac{1}{n}\sum _{i=1}^{n} x_{i}
\  , \ \
\sigma_{x}= \sqrt{  \frac{1}{n-1} \sum _{i=1} ^{n}(x_{i}-\mu_{x})^2 }
,$$
$$
\sigma_{xy}=\frac{1}{n-1}\sum _{i=1}^{n} (x_{i}-\mu_{x})(y_{i}-\mu_{y}).
$$
In theory, the correlation coefficient $\frac{\sigma_{xy}}{\sigma_x \sigma_y}$
 ranges  in the interval $[-1, 1]$, and therefore the Pearson distance  ranges in the interval $[0,2]$.

The last distance we considered is  based on the   information distance defined in \cite{Li2004}. The use of this distance  is motivated computationally since it is easily computed from FCGRs as it tracks the number of different $k$-mers for a sequence instead of the actual set. In \cite{Li2004}, for a given $k$, the information distance for two strings $x,y$ is defined as 

$$
d_{AID}(x,y)=\frac{N_{k}(x|y)+N_{k}(y|x)}{N_{k}(xy)}$$
with
$$
N_{k}(x|y)=N_{k}(xy)-N_{k}(x)
$$
where $N_{k}(x)$ is the number of different $k$-mers (possibly overlapping) which occur in $x$. We go one step further and modify this in order to avoid the creation of ``unwanted'' $k$-mers from the concatenation $xy$ of $x$ and $y$. First, we need to show how we compute $N_{k}(x)$ for a sequence $x$. For a sequence $x$, firstly, we build its FCGR$(x)=X \in \mathbb{N} ^{2^{k} \times 2^{k}}$, which is a matrix of $2^{k} \times 2^{k}$ with element values in $\mathbb{N}$. Then we unitize $X$, that is every non-zero entry becomes $1$, while zeros remain $0$.  $N_{k}(x)$ is now computed as the sum of the elements of this unitized FCGR, that is,  $ N_{k}(x) = f(X) = \text{SumOfElements}(\text{Unitize}(X))$. For two strings $x$ and $y$, with FCGRs $X$ and $Y$ respectively, we define $N_{k}(x|y)$  as:

\begin{equation}
 \label{eq:infodist_modified}
N_{k}(x|y)= f(X+Y)-N_{k}(x) 
\end{equation}

This slight modification of the information distance gives us also the desired properties of $d(x,x)=0$ and $d(x,y)=d(y,x)$ which were not satisfied before. Using  (\ref{eq:infodist_modified}),  we now define the {\it approximated information distance} (AID) as:

\begin{equation}
 \label{eq:infodist_final}
d_{AID}(x,y)= 2-\frac{f(X)+f(Y)}{f(X+Y)}
\end{equation}
where $x,y$ are the strings and $X,Y \in \mathbb{N} ^{2^{k} \times 2^{k}}$ their FCGRs, respectively. It also turns out that this distance is in fact the normalised Hamming Distance of the unitized FCGRs $X$ and $Y$.
Note that, for two sets $\mathcal X$ and $\mathcal Y$, the normalized Hamming distance is 
$\frac{|\mathcal X  \bigtriangleup \mathcal Y|}{|\mathcal X \cup \mathcal Y|} = 
	2 - \frac{|\mathcal X| + |\mathcal Y|}{|\mathcal X \cup \mathcal Y|}$
where $\bigtriangleup$ denotes the symmetric difference.

The generation of CGR images, calculation of distance matrices and creation of 2D and 3D Molecular Distance Maps with MDS were done and can be tested with the code available in \cite{gitIntraSupplemental} written in Wolfram Mathematica, version 9.  The interactive webtool ModMap, \cite{gitIntraModMap}, allows in-depth exploration of the 2D Mod Maps (Molecular Distance Maps) in this paper\footnote{When using the interactive webtool MoDMap,   clicking on a   distance  underneath a dataset will  result in plotting the  MoD Map of the  dataset computed with that distance. On any particular MoD Map,  clicking on a point will display  a window with information about the subsequence represented by that point: its NCBI accession number, scientific name of the organism it originates from,  and its CGR pattern. Clicking on the ``From here''  and ``To here'' buttons on two such selected  windows will display the distance between the corresponding genomic subsequences in the distance matrix.}. Online Supplemental Material \cite{gitIntraSupplemental} includes all distance matrices and  the  code used to produce all figures and plots in this paper. More details about the online resources can be found in Appendix B.

\section*{Analysis and Results}

For our dataset, we  use $k=9$, that is,  each  DNA sequence was represented as
a $2^9 \times 2^9$ FCGR  matrix. In practice, this means that the FCGR of a DNA sequence contains the
 full information regarding  its   $k$-mer sequence composition, for $k = 1, 2,..., 9$. The length choice of 150 kbp and value of $k = 9$  is justified by  the fact that, for a random sequence of  length 150 kbp, its CGR at resolution  $2^9 \times 2^9$  has  around  half of the  pixels black, and half white.

\begin{figure}[ht!]
\centering  
\subfigure[DSSIM distance.]{
    \includegraphics[width=0.43\linewidth]{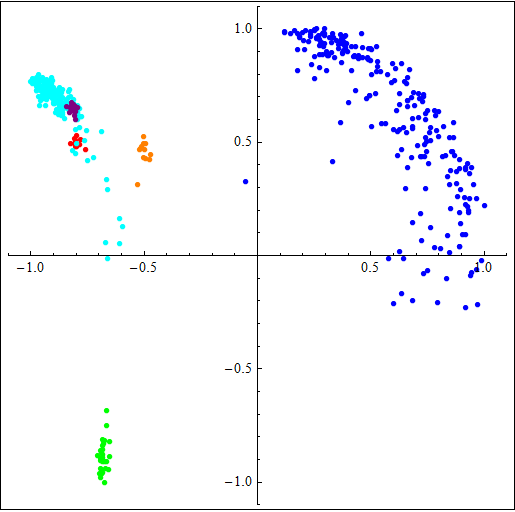}
    \label{fig:2dmds_1}
}
\subfigure[Descriptors distance.]{
    \includegraphics[width=0.43\linewidth]{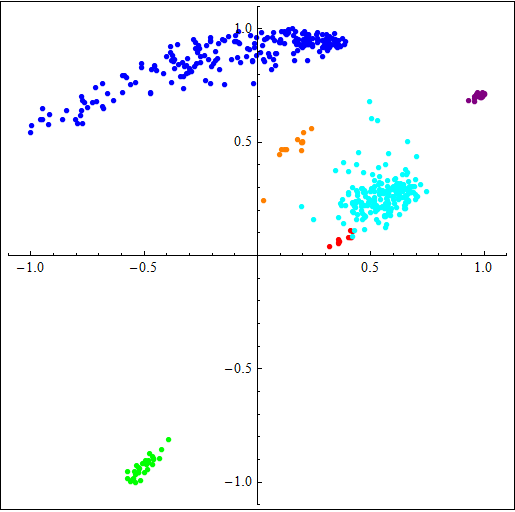}
    \label{fig:2dmds_2}
}
\\
\subfigure[Euclidean distance]{
    \includegraphics[width=0.43\linewidth]{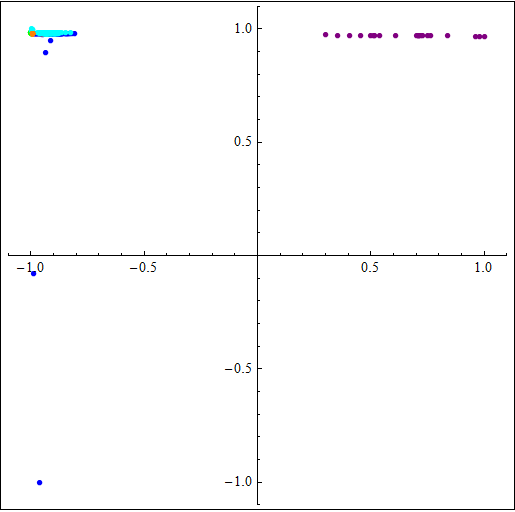}
    \label{fig:2dmds_3}
}
\subfigure[Manhattan distance]{
    \includegraphics[width=0.43\linewidth]{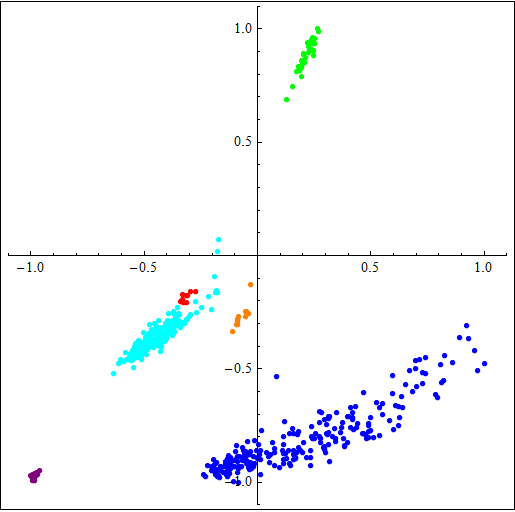}
    \label{fig:2dmds_4}
}
\\
\subfigure[Pearson distance]{
    \includegraphics[width=0.43\linewidth]{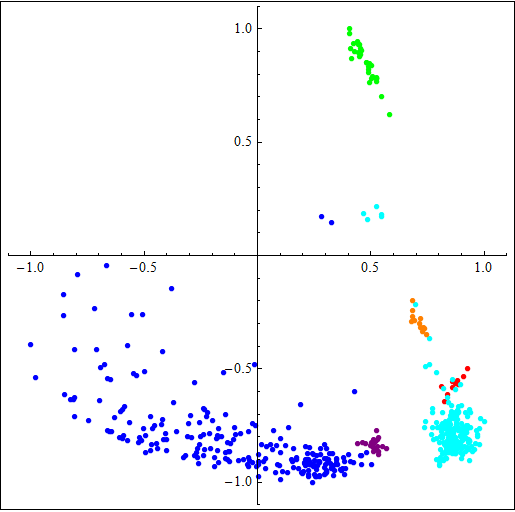}
    \label{fig:2dmds_5}
}
\subfigure[Approx. inform. distance]{
    \includegraphics[width=0.43\linewidth]{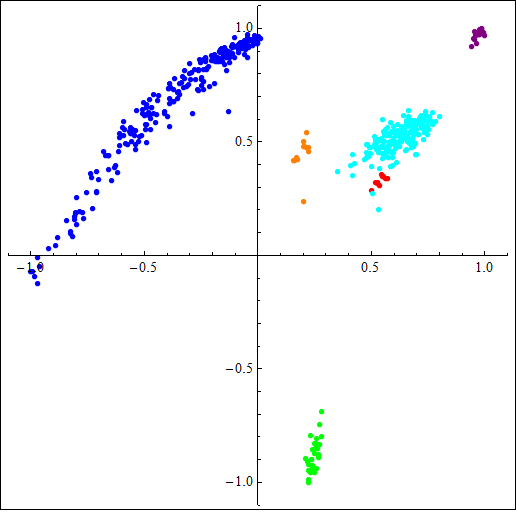}
    \label{fig:2dmds_6}
}
%
\caption{Two-dimensional Molecular Distance Maps  of DNA genomic sequences from all six organisms  in the dataset, obtained using DSSIM, descriptor, Euclidean, Manhattan, Pearson and aproximated information  distance, respectively. Each point corresponds to a 150 kbp   genomic sequence from  \textit{H.~sapiens} (blue), \textit{E.~coli} (green), \textit{S.~cerevisiae} (red), \textit{A.~thaliana} (turqoise), \textit{P.~falciparum}  (magenta),  and \textit{P.~furiosus} (orange).}
\label{fig:all_2dmds}
\end{figure}

\begin{figure}[ht!]
\centering
\subfigure[DSSIM distance.]{
    \includegraphics[width=0.43\linewidth]{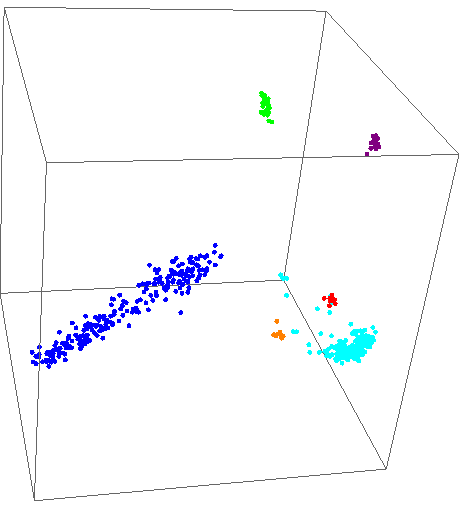}
    \label{fig:3dmds_1}
}
\subfigure[Descriptors distance.]{
    \includegraphics[width=0.43\linewidth]{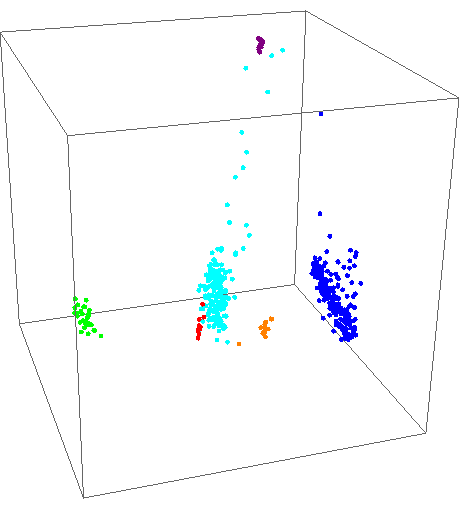}
    \label{fig:3dmds_2}
}
\\
\subfigure[Euclidean distance]{
    \includegraphics[width=0.43\linewidth]{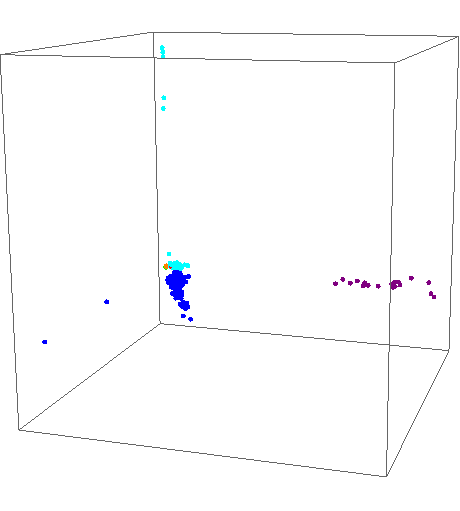}
    \label{fig:3dmds_3}
}
\subfigure[Manhattan distance]{
    \includegraphics[width=0.43\linewidth]{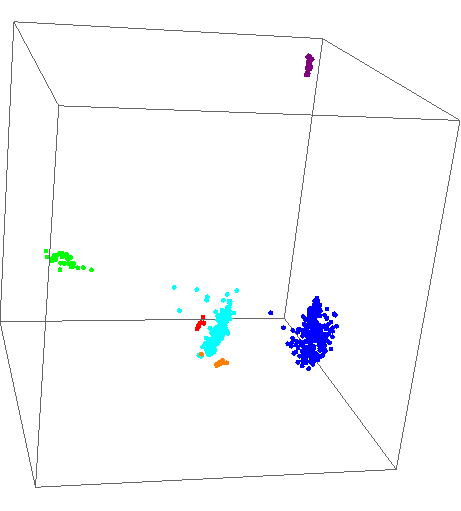}
    \label{fig:3dmds_4}
}
\\
\subfigure[Pearson distance]{
    \includegraphics[width=0.43\linewidth]{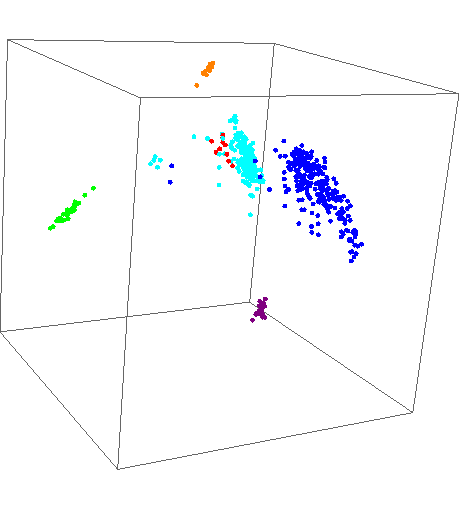}
    \label{fig:3dmds_5}
}
\subfigure[Approx. inform. distance]{
    \includegraphics[width=0.43\linewidth]{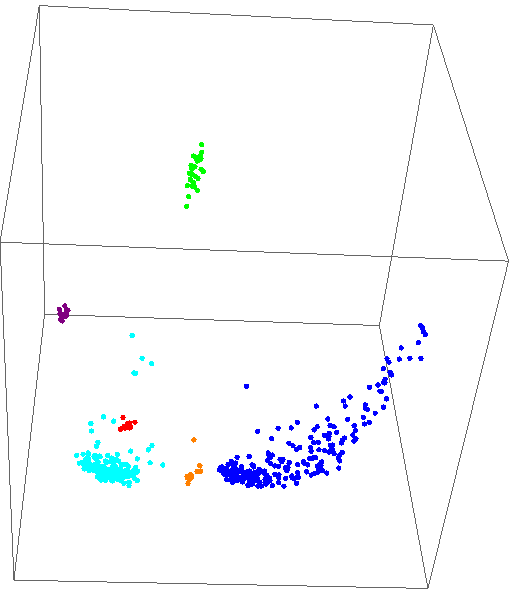}
    \label{fig:3dmds_6}
}
\caption{Three-dimensional Molecular Distance Maps  of  genomic DNA sequences from all six organisms  in the dataset, obtained using DSSIM, descriptor, Euclidean, Manhattan, Pearson and approximated information distance, respectively. Each point corresponds to a 150 kbp   genomic sequence from  \textit{H.~sapiens} (blue), \textit{E.~coli} (green), \textit{S.~cerevisiae} (red), \textit{A.~thaliana} (turqoise), \textit{P.~falciparum}  (magenta),  and \textit{P.~furiosus} (orange).}
\label{fig:all_3dmds}
\end{figure}

Figure \ref{fig:all_2dmds} depicts two-dimensional Molecular Distance Maps for the over five hundred  DNA sequences in our  dataset, computed using the DSSIM distance, descriptor distance, Euclidean distance, Manhattan distance, Pearson distance  and approximated information distance, respectively. Figure~\ref{fig:all_3dmds}  depicts the corresponding three-dimensional Molecular Distance Maps for the same dataset. The projection of each three-dimensional map is chosen by hand in order to visually separate clusters of points which appear to be overlapping in the two-dimensional maps, as discussed below.

We note that MDS is not a clustering method, as the clusters are defined beforehand by the coloring scheme used (blue for {\it H.~sapiens}, green for {\it E.~coli}, and so on). MDS simply tries to display visually the  interrelationships between the given items, based on the pairwise distances in the distance matrix which is its input. Note also that an increase in dimensionality from  2 to 3  can lead to a better cluster visualization. For example,  if we compare the two-dimensional and the three-dimensional Molecular Distance Maps obtained using DSSIM, we see that points that appeared to be erroneously mixed with each other in the two-dimensional map, Figure \ref{fig:all_2dmds}(a),
 ({\it S. cerevisiae}  and {\it P. falciparum}  sequences mixed in with  {\it A. thaliana}  sequences) were in fact clearly separated from each other in Figure 
 \ref{fig:all_3dmds}(a),  the three-dimensional version  of the Molecular Distance Map. 

\begin{figure}[ht!]
\centering
\subfigure[DSSIM distance.]{
    \includegraphics[width=0.43\linewidth]{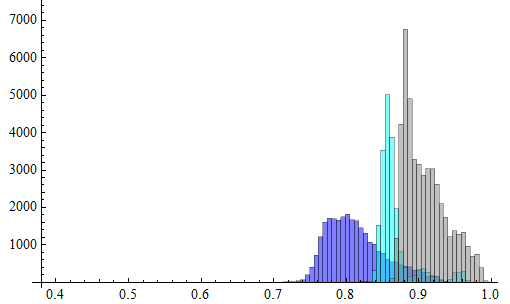}
    \label{fig:hist1_1}
}
\subfigure[Descriptors distance.]{
    \includegraphics[width=0.43\linewidth]{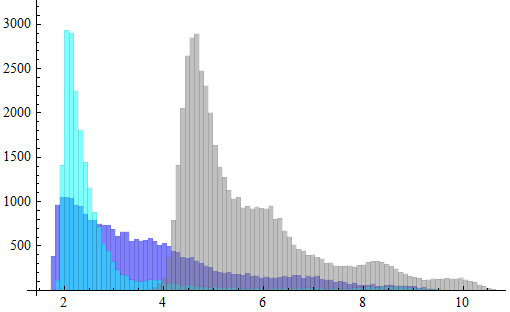}
    \label{fig:hist1_2}
}
\\
\subfigure[Euclidean distance]{
    \includegraphics[width=0.43\linewidth]{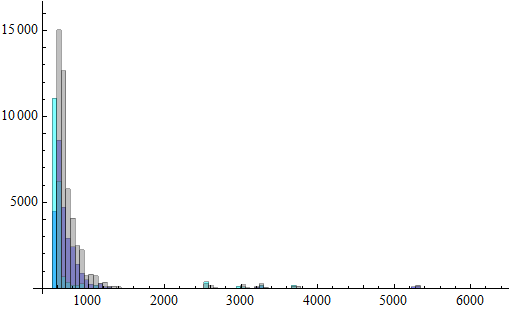}
    \label{fig:hist1_3}
}
\subfigure[Manhattan distance]{
    \includegraphics[width=0.43\linewidth]{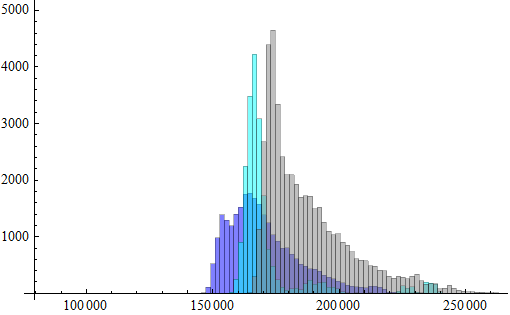}
    \label{fig:hist1_4}
}
\\
\subfigure[Pearson distance]{
    \includegraphics[width=0.43\linewidth]{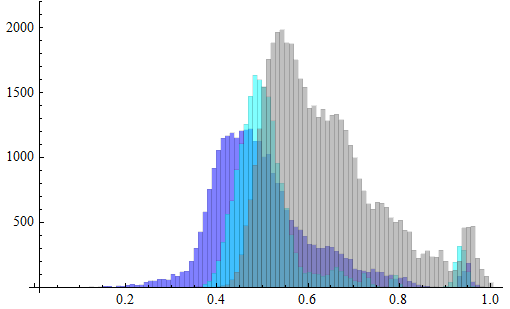}
    \label{fig:hist1_5}
}
\subfigure[Approx. inform. distance]{
    \includegraphics[width=0.43\linewidth]{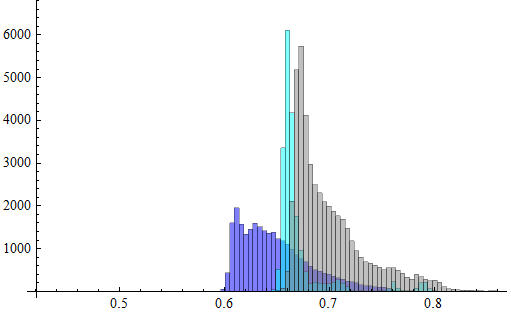}
    \label{fig:hist1_6}
}
\subfigure
{
    \includegraphics[width=0.43\linewidth]{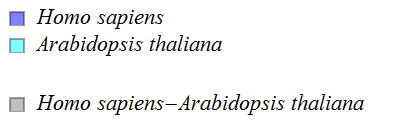}
    \label{fig:hist1_7}
}
\caption{Histograms of pairwise  intragenomic and intergenomic distances  among the  DNA sequences from \textit{H.~sapiens} and \textit{A.~thaliana}.}
\label{fig:all_hist1}
\end{figure}

Figure \ref{fig:all_hist1} displays the histograms of the pairwise  intragenomic distances (dark blue  and turqoise) and  intergenomic distances (grey) of DNA sequences from {\it H.~sapiens}  and {\it A.~thaliana}, obtained using each of the six distances.  As noted, some distances seem to perform better than others. Visually, the poorest performer for these two sets of sequences (from {\it H.~sapiens} and {\it A.~thaliana})  seems to be the Euclidean distance wherein the intragenomic distances are as high as intergenomic distances,  and no separation is visible. In contrast, DSSIM  gives -- for the same data -- intergenomic distances that  are overall much higher than intragenomic distances,  resulting in a clear classification of DNA sequences into the species they belong to. 

Table  \ref{6tables} displays the mean and standard deviation of distances between clusters $C_i$ and $C_j$,  $1\leq i, j \leq 6$, 
where a cluster $C_\ell$  is defined as  the set of all genomic sequences  from the genome of organism $\ell$, as labelled in Table \ref{table1}.  In each subtable, the diagonals represent the means and standard deviation for intragenomic distances, while the other entries are all intergenomic distances.
From this table we see that
 for DSSIM, Manhattan and   approximated information distance, the maximum of all the averages  of intragenomic distances in this dataset  is strictly smaller than the minimum of  all the  averages of  intergenomic distances. For  the descriptor distance and Pearson distance the previous statement does not hold but, for each pair of organisms, the  two averages of intragenomic distances (e.g., human-human and plant-plant) are both lower than the average of the intergenomic distances (human-plant). For the Euclidean distance, none of the previous statements holds: For example,   the average of the plant-plant  intragenomic distances (element 4-4 in the Euclidean distance   subtable of Table \ref{6tables}) intragenomic distances  is 723, which is  larger than  672, the average of the yeast-plant intergenomic distances (element 3-4 in the Euclidean distance   subtable of Table \ref{6tables}).  The complete histograms of all pairwise comparisons $C_i - C_j$ can be found in   Appendix~C.

\begin{table}[H]
\begin{scriptsize}
\begin{tabular}{|@{\hspace{1mm}}L{0.1cm}@{\hspace{1mm}}| @{\hspace{1mm}}C{0.8cm}@{\hspace{1mm}}|@{\hspace{1mm}}C{0.8cm}@{\hspace{1mm}}|@{\hspace{1mm}}C{0.8cm}@{\hspace{1mm}}|@{\hspace{1mm}}C{1cm}@{\hspace{1mm}}|@{\hspace{1mm}}C{1cm}@{\hspace{1mm}}|@{\hspace{1mm}}C{1cm}@{\hspace{1mm}}|}
\hline
- & 1 & 2 & 3 & 4 & 5 & 6\\ \hline
1& $0.81\pm0.04$ & $0.99\pm0.01$ & $0.92\pm0.02$ & $0.91\pm0.03$ & $0.92\pm0.03$ & $0.91\pm0.02$ \\ \hline
2& - & $0.85\pm0.01$ & $0.97\pm0.01$ & $0.99\pm0.01$ & $0.99\pm0.01$ & $0.99\pm0.$ \\ \hline
3& - & - & $0.87\pm0.01$ & $0.89\pm0.02$ & $0.91\pm0.$ & $0.91\pm0.01$ \\ \hline
4& - & - & - & $0.87\pm0.03$ & $0.9\pm0.02$ & $0.91\pm0.01$ \\ \hline
5& - & - & - & - & $0.74\pm0.01$ & $0.94\pm0.$ \\ \hline
6& \multicolumn{4}{l}{\textbf{DSSIM}}  &  & $0.83\pm0.01$ \\ \hline  \multicolumn{6}{l}{} \\ \hline 
1&  $3.76\pm1.69$ & $9.74\pm0.66$ & $5.92\pm1.14$ & $5.71\pm1.41$ & $9.33\pm1.23$ & $5.44\pm0.92$ \\ \hline
2& - & $2.5\pm0.28$ & $8.05\pm0.39$ & $9.1\pm0.55$ & $12.67\pm0.19$ & $9.38\pm0.41$ \\ \hline
3& - & - & $2.12\pm0.08$ & $3.42\pm1.05$ & $9.48\pm0.31$ & $4.6\pm0.09$ \\ \hline
4& - & - & - & $2.75\pm1.33$ & $8.23\pm0.94$ & $4.94\pm0.76$ \\ \hline
5& - & - & - & - & $1.53\pm0.14$ & $9.99\pm0.28$ \\ \hline
6& \multicolumn{4}{l}{\textbf{Descriptors}} &  & $2.4\pm0.32$ \\ \hline \multicolumn{6}{l}{} \\ \hline 
1& $756\pm498$ & $856\pm349$ & $756\pm361$ & $818\pm514$ & $3914\pm510$ & $812\pm356$ \\ \hline
2& - & $558\pm5$ & $674\pm17$ & $802\pm366$ & $4102\pm466$ & $696\pm18$ \\ \hline
3& - & - & $564\pm11$ & $672\pm383$ & $3964\pm472$ & $633\pm20$ \\ \hline
4& - & - & - & $723\pm535$ & $3923\pm506$ & $748\pm372$ \\ \hline
5& - & - & - & - & $999\pm276$ & $4085\pm468$ \\ \hline
6& \multicolumn{4}{l}{\textbf{Euclidean}}  &  & $585\pm 24$ \\[0ex] \hline \multicolumn{6}{l}{} \\ \hline 
1& $171\pm15$ & $222\pm5$ & $189\pm13$ & $188\pm17$ & $213\pm20$ & $191\pm9$ \\[0ex]\hline
2& - & $175\pm2$ & $209\pm4$ & $219\pm8$ & $252\pm4$ & $218\pm3$ \\[0ex] \hline
3& - & - & $171\pm2$ & $177\pm10$ & $206\pm2$ & $184\pm2$ \\[0ex] \hline
4& - & - & - & $172\pm16$ & $200\pm11$ & $188\pm9$ \\ [0ex] \hline
5& - & - & - & - & $105\pm3$ & $224\pm2$ \\[01ex] \hline
6& \multicolumn{4}{l}{ \textbf{Manhattan (in thousands) }}&  & $167\pm3$ \\[2ex] \hline \multicolumn{6}{l}{} \\ \hline
1& $0.5\pm0.12$ & $0.97\pm0.02$ & $0.69\pm0.1$ & $0.64\pm0.12$ & $0.65\pm0.09$ & $0.81\pm0.06$ \\ \hline
2& - & $0.71\pm0.02$ & $0.93\pm0.02$ & $0.96\pm0.02$ & $0.98\pm0.01$ & $0.99\pm0.02$ \\ \hline
3& - & - & $0.6\pm0.02$ & $0.6\pm0.07$ & $0.71\pm0.03$ & $0.75\pm0.02$ \\ \hline
4& - & - & - & $0.53\pm0.11$ & $0.63\pm0.09$ & $0.76\pm0.04$ \\ \hline
5& - & - & - & - & $0.02\pm0.01$ & $0.94\pm0.01$ \\ \hline
6& \multicolumn{4}{l}{\textbf{Pearson}}  &  & $0.64\pm0.03$ \\ \hline \multicolumn{6}{l}{} \\ \hline
1& $0.65\pm0.03$ & $0.78\pm0.01$ & $0.7\pm0.03$ & $0.7\pm0.03$ & $0.76\pm0.04$ & $0.69\pm0.02$ \\ \hline
2& - & $0.67\pm0.$ & $0.75\pm0.01$ & $0.77\pm0.02$ & $0.85\pm0.01$ & $0.77\pm0.01$ \\ \hline
3& - & - & $0.67\pm0.01$ & $0.68\pm0.02$ & $0.74\pm0.$ & $0.69\pm0.$ \\ \hline
4& - & - & - & $0.67\pm0.03$ & $0.73\pm0.02$ & $0.69\pm0.02$ \\ \hline
5& - & - & - & - & $0.64\pm0.01$ & $0.76\pm0.01$ \\ \hline
6& \multicolumn{4}{l}{ \textbf{Approx. Information}}  & & $0.65\pm0.01$ \\ \hline
\end{tabular}
\end{scriptsize}
\vspace{3mm}
\caption{Mean and standard deviation of distances between clusters $C_i - C_j$ for $i,j = 1, ...,  6$.}
\label{6tables}
\end{table}

\subsection*{Quality Measures for   Distances}\label{sec:quality}

In this section we present three  quality measures  that each evaluates the quality of the six distances considered. In the  data mining literature a wide range of quality measures for clusterings has been defined; see for example \cite{tan2007,zhao2004empirical}.
Most of these methods are designed to assess the quality of different automated clustering methods while using the same distance. Our set-up is different, as we use different distances while the clustering is fixed and given by the initial colour-coding of the sequence-representing points. Thus, we have to use other approaches to  compare the distances we analyze. In  particular, as the six distances have  different ranges, we have to use assessment methods which are invariant to the scale of the distance.

The ``ground-truth'' that we use  as a basis for our distance assessment is the fact that the ``ideal'' clustering of DNA sequences and the points that represent them is known: sequences from the same organism should be close to one another and far from sequences originating from other organisms. (This assumption is justified -- for this dataset -- as the six organisms considered are very different from one another, belonging to different kingdoms of life.)
Thus, an optimal distance  should yield a relatively small distance between two FCGRs which were generated from the DNA sequences originating from  the same organism, and relatively high distances between two FCGR originating from DNA sequences coming from different organisms.

In order to assess each of the six distances quantitatively, we computed three quality  measures which rate different features of a distance:

\begin{itemize}
\item the correlation to an idealized cluster distance

\item the silhouette cluster accuracy
\item the relative overlap between the intragenomic and intergenomic distance histograms.
\end{itemize}
Let us stress that all three quality measures of the six distances are based on the distance matrices which we computed and not on their MDS plots.
We will define the three quality measures such that their expected values range in the interval $[0,1]$ where higher values correspond to better performance.

Let us first describe the three quality measures informally. An idealized distance  is a distance that would be able  to differentiate DNA sequences by species, that is, a distance $\delta$ for which $\delta(x,y) = 0$ if $x$ and $y$ are sequences from the same species and $\delta(x,y) = 1$ otherwise.
The  first quality measure, the \emph{correlation to an idealized cluster distance},  measures how well a distance is linearly correlated to the idealized distance $\delta$.
The  second quality measure, \emph{silhouette cluster accuracy}, is the percentage of points that are best embedded in the cluster they belong to.
The third quality measure quantifies the ``visual overlap'' between the intragenomic and intergenomic distance histograms. Given our dataset, it is reasonable to expect that a good distance gives a low value if applied to FCGRs of  genomic sequences of the same organism, and a high value when applied to  FCGRs of genomic sequences from two different organisms, thus separating the histograms of intragenomic distances from that of intergenomic distances. This is illustrated by the histograms in Figure~\ref{fig:all_hist1}, where a high overlap between the graph of intragenomic distances (dark blue and turquoise) and the graphs of intergenomic distances (grey) is an indication of a poorly performing distance. 
In a theoretically optimal situation, there would exist a value $c$ such that all distances that are smaller than $c$ are intragenomic distances and all distances that are larger than $c$ are intergenomic distances. This can usually not be expected from real data, but a low overlap between histograms is nevertheless indicative of a ``good'' distance.

In order to formally define the three quality measures, we consider a dataset $V$ which is partitioned into $p$ non-overlapping clusters $C_1,\ldots, C_p$ for which a distance  $d_\alpha\colon V\times V \to \mathbb{R}_{\ge 0}$ exists.
The cardinalities of the sets are $ |V| = m$ and $|C_i| = m_i$ for $i = 1,\ldots, p$.
In our analysis, $p=6$ and  $C_1$ contains all FCGRs generated from  genomic DNA sequences from  {\it H.~sapiens}, $C_2$ contains all FCGRs generated from genomic sequences of  {\it E.coli}, and so on, according to the order in Table \ref{table1}. The distance $d_\alpha$ is one of the six distances $\alpha\in\{$DSSIM, D, E, M, P, AID$\}$.

The {\it correlation to an idealized cluster distance} is computed as follows. 
We define the {\it idealized cluster distance} as a function (or matrix) $\delta \colon V\times V \to \{0,1\}$ such that $\delta(x,y) = 0$ if and only if $x$ and $y$ belong to the same cluster, and $\delta(x,y) = 1$ otherwise. 
Because we can view $d_\alpha$ and $\delta$ as discrete, symmetric functions which have the same domain, we can compute their correlation coefficient.
We define the correlation of $\delta$ to $d_\alpha$ to be the Pearson correlation of $\delta$ and $d_\alpha$.
More precisely, the upper triangular part  of the matrix  corresponding to a distance $d_\alpha$ is interpreted as a vector $(x_1,\ldots, x_n)$ and compared with the corresponding values $(y_1,\ldots,y_n)$ given by $\delta$.
We obtain the $\delta$-correlation as

$$
	\mathcal D_\alpha = \frac{\sigma_{xy}}{\sigma_x\sigma_y}.
$$
The correlation ranges in  the interval $[-1,1]$: a value of $1$ means that $d_\alpha$ and $\delta$ are linearly correlated, and a value of $0$ means that they are unrelated.
In other words, if the value obtained by measuring the {\it correlation} of a given distance {\it to the idealized cluster distance}  is close to $1$,  this means that the given  distance  is closer to the idealized cluster distance, and hence, performs well.
Note that negative values  for this measure are not expected as this would imply that $d_\alpha$ and $\delta$ were negatively related ($d_\alpha$ would perform worse than a matrix containing random entries).

The {\it silhouette cluster accuracy} is based on the  {\em silhouette coefficient}, defined in \cite{silhouette}, as a measure that determines how well a single point is embedded in the cluster to which it belongs.
For a point $x$ from cluster $C_i$ we define $a_x$ as the average distance of this point to all other points in $C_i$, that is, 

$$
	a_x = \frac1{m_i-1}\sum_{y\in C_i,y\neq x} d_{\alpha}(x,y),
$$

and we define $b_x$ as the minimum over the average distances of $x$ to all points of a different cluster

$$
	b_x = \min_{j=1,j\neq i}^K \left\{ \frac1{m_j}\sum_{y\in C_j} d_{\alpha}(x,y) \right\}.
$$

The silhouette coefficient of $x$ is defined as 

$$
	\mathcal S_\alpha(x) = \frac{b_x - a_x}{\max\{a_x,b_x\}}.
$$

If a point $x$ has a silhouette coefficient $\mathcal S_\alpha(x) \le 0$, then $x$ is at least as close to a cluster to which it does not belong than to its own cluster.
The {\it silhouette cluster accuracy} $\mathcal A_\alpha$ denotes the percentage of points with a silhouette coefficient greater than $0$, that is the percentage of points which are well-embedded in their own cluster,

$$
	\mathcal A_\alpha = \frac{\left| \{ x\in V \mid \mathcal S_\alpha(x) > 0 \} \right|}{m}.
$$
Obviously, the silhouette cluster accuracy ranges in $[0,1]$ with a high accuracy being desirable.

For assessing the \emph{relative overlap} of the histograms, 
consider any two clusters $C_i$ and $C_j$ with $i\neq j$ (for example, $C_1$ is the {\it H.~sapiens} cluster and $C_4$ the {\it A.~thaliana} cluster).
We compare the two sets of  intragenomic distances $C_i$--$C_i$ and $C_j$--$C_j$ with the set of intergenomic distances $C_i$--$C_j$.
For a distance $d_\alpha$, we divide the range from $\min (d_{\alpha})$ to the maximum distance $\max (d_{\alpha})$ in this dataset into $100$ bins of size $r = \frac{\max (d_{\alpha}) - \min (d_{\alpha})}{100}$ and count the distances which fall into this bin: $c_{i,i}[\ell]$ denotes bin $\ell$ containing distances from $C_i$--$C_i$ and $c_{i,j}[\ell]$ denotes bin $i$ containing distances from $C_i$--$C_j$.
For $\ell = 1,\ldots,100$ we let 

\begin{align*}
	c_{i',j'}[\ell] = | \{&
		\{x,y\}\mid x\in C_{i'},y\in C_{j'} \mbox{ and }  x \neq y \\
 &\mbox{and } 
		(\ell-1)\cdot r < d_\alpha(x,y) \le \ell\cdot r
		\} |.
\end{align*}

By $s_{i',j'}$ we denote the sum over all $c_{i',j'}$-bins $s_{i',j'} = \sum_{\ell=1}^{100} c_{i',j'}[\ell]$.
We define the relative overlap $\mathcal{O}_{\alpha}(i,j)$ of $C_i$--$C_i$ (intragenomic distances) with $C_i$--$C_j$ (intergenomic distances) as
$$
	\mathcal{O}_{\alpha}(i,j) = 
	\frac{\max\{s_{i,i}, s_{i,j} \}}
		{\min\{s_{i,i}, s_{i,j} \}} \cdot 
	\frac{\sum_{i=1}^{100} \min\{c_{i,i},c_{i,j}\}}
		{\sum_{i=1}^{100} \max\{c_{i,i},c_{i,j}\}}. \\
$$
The relative overlap $\mathcal{O}_{\alpha}(j,i)$ of $C_j$--$C_j$ with $C_i$--$C_j$  is defined analogously; note that $\mathcal{O}_{\alpha}(i,j) \neq \mathcal{O}_{\alpha}(j,i)$ in general.
The overlap is normalized to the range  $[0,1]$ where $0$ means no overlap of elements of bins between intra- and intergenomic distances, and $1$ means that one of the histograms  completely ``covers'' the other.
Also note that we are not interested in the overlap of $C_i$--$C_i$ with $C_j$--$C_j$ as both sets of distances are intragenomic distances.

Since we intend to define the a quality  measure where a value close to $1$ should represent a small overlap, we will use $1-\mathcal{O}_{\alpha}(i,j)$ as relative overlap.
Furthermore, we combine these  quantities for all possible pairs of clusters $C_i$ and $C_j$, obtaining {\it the relative overlap} as:

$$
	\mathcal O_\alpha = 1 - \frac{1}{p(p-1)}
	\sum_{i=1}^p\sum_{j=1,i\neq j}^p \mathcal O_\alpha(i,j).
$$

For example, in Figure~\ref{fig:all_hist1}, for each of the considered distance, the  dark blue  histograms depict the $C_1 - C_1$  ({\it H.~sapiens} -- {\it H.~sapiens}) intragenomic distances,
the  turquoise histograms the $C_4 - C_4$  ({\it A.~thaliana} -- {\it A.~thaliana}) intragenomic distances, and grey  histograms the  $C_1 - C_4$ ({\it H.~sapiens} -- {\it A.~thaliana}) intergenomic distances. As seen from this figure, the descriptor distance appears to visually perform best at separating the two intragenomic distance histograms from the intergenomic histogram, while the Euclidean distance has the weakest performance. The relative overlap attempts to quantify this by computing the  overlaps  of each of the two  pairs of histograms (dark blue with grey and turquoise with grey). Note that  small visual histogram overlaps  will result in a high numerical {\it relative overlap}, and is indicative of a better performing distance. 
 
\subsection*{Distance Comparison Results}

The results of  comparing the six distances we analyzed, using the three quality measures,  are listed in Table~\ref{tab:quality}. Recall that all quality measures have an expected range of $[0,1]$ where larger values imply better performance.

\begin{table}[ht]
\centering
\begin{tabular}{|l|r|r|r|r|c|r|c|}
\hline
& \multicolumn{1}{c|}{$\mathcal D_\alpha$} 
& \multicolumn{1}{c|}{$\mathcal A_\alpha$}
& \multicolumn{1}{c|}{$\mathcal O_\alpha$}
& \multicolumn{1}{c|}{$z$-score sum} 
& \multicolumn{1}{c|}{ Rank}\\
\hline
DSSIM          &$0.627$ & $1.000$ & $0.965$ &   $1.895$   & $2$nd  \\ \hline 
Descriptors  &$0.639$ & $0.976$ & $0.988$ &    $2.509$   & $1$st  \\\hline 
Euclidean     &$0.231$ & $0.325$ & $0.907$ &   $-4.831$  & $6$th  \\ \hline
Manhattan   &$0.527$ & $1.000$ & $0.951$ &    $0.84$    & $3$rd  \\ \hline 
Pearson       &$0.536$ & $0.980$ & $0.888$ &   $-0.875$  & $5$th  \\ \hline
Approx. Inf.   &$0.527$ & $1.000$ & $0.937$ & $0.462$   & $4$th  \\\hline
\end{tabular}
\vspace{3mm}
\caption{Summary of quality measures for the performances of six distances (DSSIM, descriptors, Euclidean, Manhattan, Pearson, approximated information distance)  on a dataset of 508 genomic DNA sequences  taken from organisms from each kingdom of life. $\mathcal{D}_\alpha$ is the correlation to an idealized cluster,  $\mathcal{A}_\alpha$  the silhouette cluster accuracy,  and  $\mathcal{O}_\alpha$  the relative overlap. Higher is better.}
\label{tab:quality}
\end{table}

To compare each distance relative to all the other distances, we further compute for each quality measure (each column)  
the {\em standard scores}  ($z$-scores) of  each distance $d_{\alpha}$, where
$\alpha\in\{$DSSIM, D, E, M, P, AID$\}$,
  as 
$z(d_{\alpha}) = \frac{d_{\alpha} - \mu}{\sigma}$ where $\mu$ is the mean and $\sigma$ is the deviation of all six $d_{\alpha}$  for that particular quality measure (column).
A positive value of the standard score will  mean that a distance  performs above average (in this category) and a  negative value that it performs below average.

Finally, we compute the sum of the $z$-scores for each quality measure as seen in  Table~\ref{tab:quality}. Note that the total  of $z$-scores for a distance represents the performance of that distance relative to the other distances, and indicates its relative ranking.

The conclusion of this analysis is that the  best performing distances  are the descriptor distance and DSSIM. 
Manhattan, Pearson, and approximate information distance perform well in some categories but not so well in other categories. 
For this dataset  and value of $k$,  the Euclidean distance had the weakest performance in all measured categories, which confirms the visual assessment of the MDS plots  obtained by using the  Euclidean distance, as seen in Figure \ref{fig:all_2dmds} and Figure \ref{fig:all_3dmds}.

It is worth noting that the two distances which perform best (DSSIM and descriptor) treat FCGR matrices as two-dimensional maps in which the local arrangement of the cells (matrix entries) influences the computed distance, whereas the other distances treat the FCGR matrices as linear vectors.
This suggests that the organization of the $k$-mer tallies (in this paper $k=9$) of a DNA sequence as an FCGR matrix, rather than a simple vector, reveals  structural properties of the DNA sequence that could be utilized in order to identify and classify genomic DNA sequences. 

\section*{Discussion and Conclusions}

In this study we test the hypothesis that  CGR-based genomic signatures of genomic DNA sequences are indeed species and genome-specific. With this goal in mind we analyze over five hundred 
150 kbp DNA genomic sequences originating from organisms representing each of the kingdoms of life. 
Our quantitative comparison of six different distances suggests that several other distances outperform the Euclidean distance, which has been until now almost exclusively used in such studies. Our preliminary results show that two of these distances, DSSIM and descriptor distance (introduced here) when applied to CGR-based genomic signatures, have indeed the ability to differentiate between DNA sequences coming from different species. This indicates that the $k$-mer sequence composition (where $k= 1, 2, ..., 9$) of genomic sequences contains taxonomic information which could potentially aid in the identification, comparison and classification of species based on molecular evidence. The  two-dimensional  and three-dimensional Molecular Distance Maps we obtain, which visualize the simultaneous intragenomic and intergenomic interrelationships among the  sequences in our dataset, show this method's potential. 

Further analysis is needed to explore this method's potential to the analysis of closely related species.
As a preliminary experiment, we applied it to  \textit{H. sapiens} chromosome $21$ (NC$\_000021.8$), which yields $234$ fragments, and \textit{P. troglodytes} chromosome Y (NC$\_006492.3$) which yields $168$ sequences,  also 150  kbp long.

\begin{figure}[ht]
\centering
\subfigure[DSSIM distance.]{
    \includegraphics[width=0.43\linewidth]{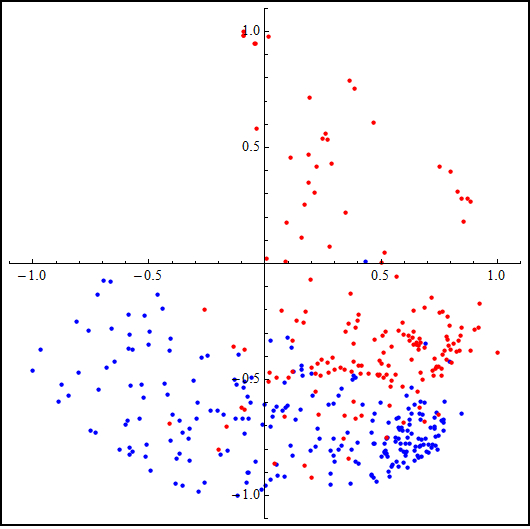}
    \label{fig:humchimp1}
}
\subfigure[Descriptors distance.]{
    \includegraphics[width=0.43\linewidth]{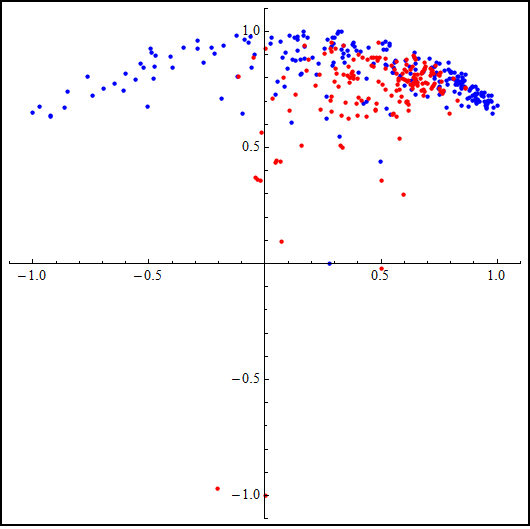}
    \label{fig:humchimp2}
}\\
\subfigure[Euclidean distance.]{
    \includegraphics[width=0.43\linewidth]{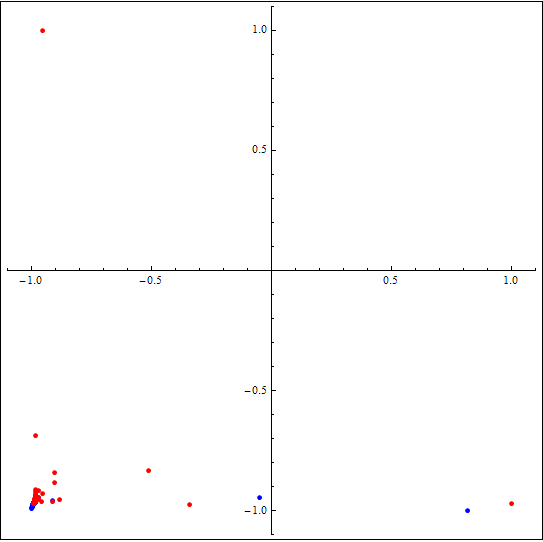}
    \label{fig:humchimp3}
}
\subfigure[Manhattan distance.]{
    \includegraphics[width=0.43\linewidth]{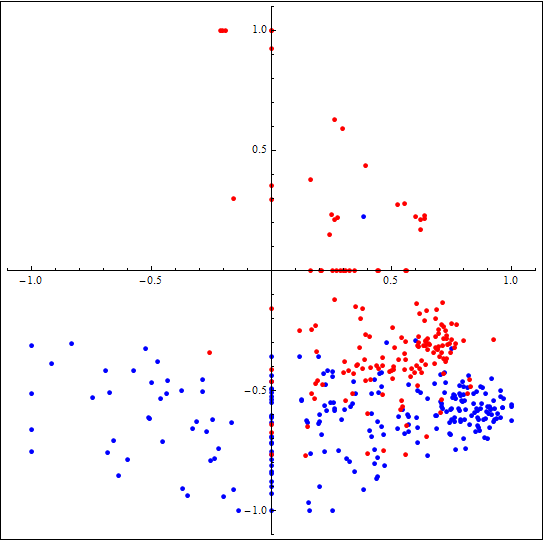}
    \label{fig:humchimp4}
}\\
\subfigure[Pearson distance.]{
    \includegraphics[width=0.43\linewidth]{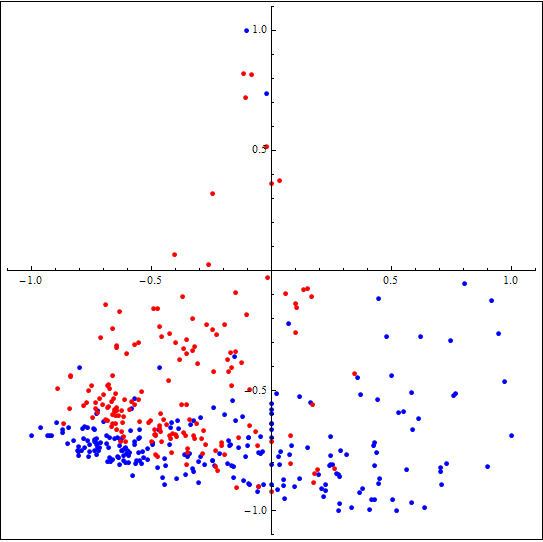}
    \label{fig:humchimp5}
}
\subfigure[Approx. inform. distance.]{
    \includegraphics[width=0.43\linewidth]{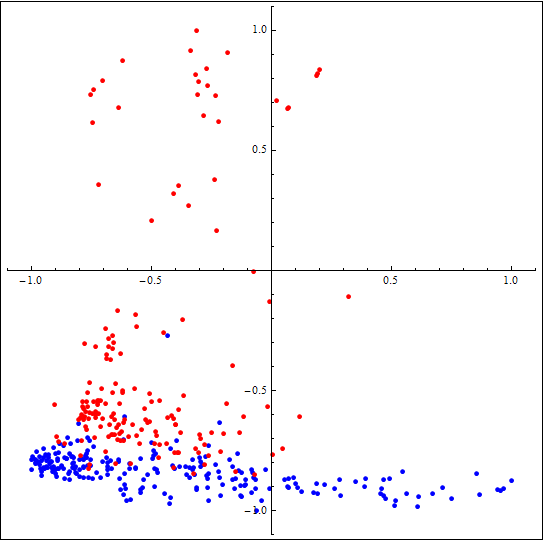}
    \label{fig:humchimp6}
}
\caption{Two-dimensional Molecular Distance Maps  of  150~kbp genomic DNA sequences from \textit{H.~sapiens} (blue), \textit{P.~troglodytes} (red) using the six distances.}
\label{fig:all_humchimp_2d}
\end{figure}

\begin{figure}[ht]
\centering
\subfigure[DSSIM distance.]{
    \includegraphics[width=0.43\linewidth]{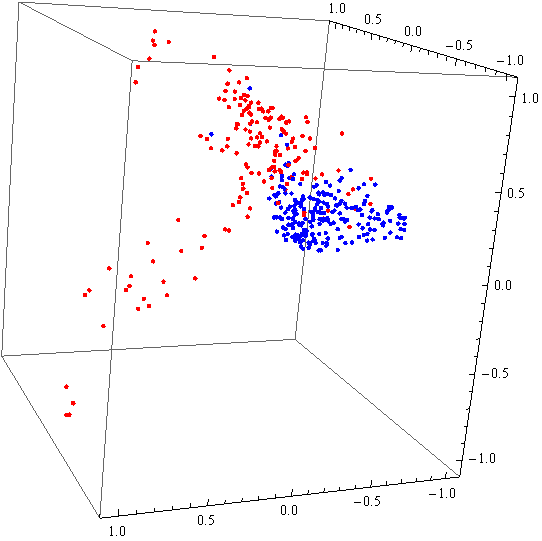}
    \label{fig:humchimp7}
}
\subfigure[Descriptors distance.]{
    \includegraphics[width=0.43\linewidth]{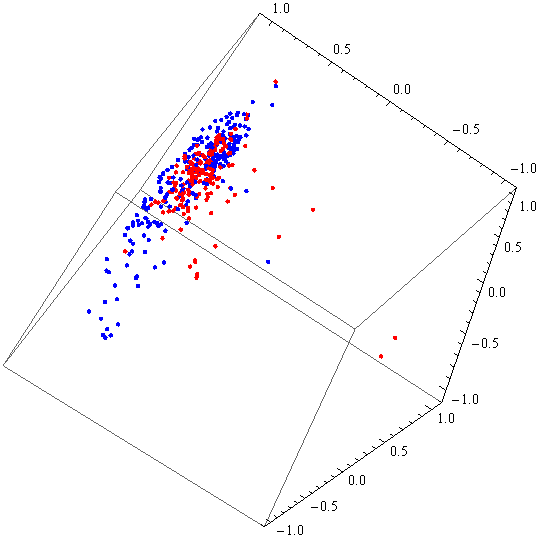}
    \label{fig:humchimp8}
}\\
\subfigure[Euclidean distance.]{
    \includegraphics[width=0.43\linewidth]{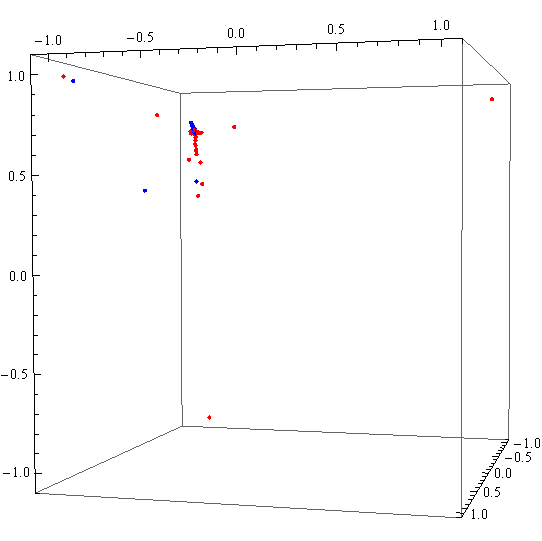}
    \label{fig:humchimp9}
}
\subfigure[Manhattan distance.]{
    \includegraphics[width=0.43\linewidth]{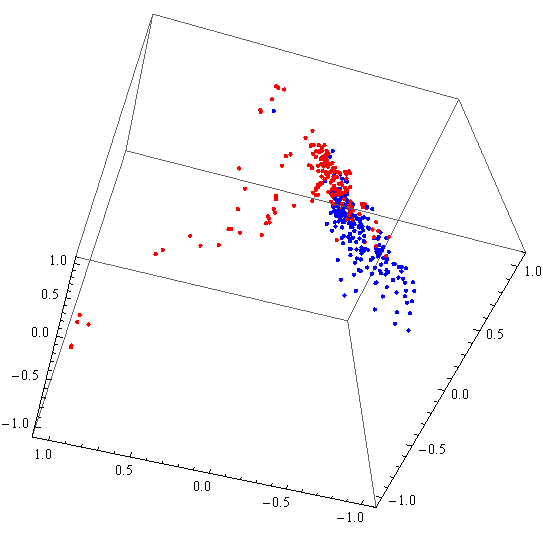}
    \label{fig:humchimp10}
}\\
\subfigure[Pearson distance.]{
    \includegraphics[width=0.43\linewidth]{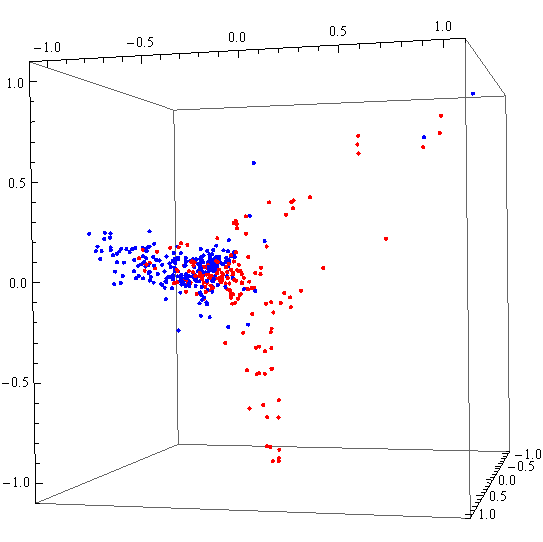}
    \label{fig:humchimp11}
}
\subfigure[Approx. inform. distance.]{
    \includegraphics[width=0.43\linewidth]{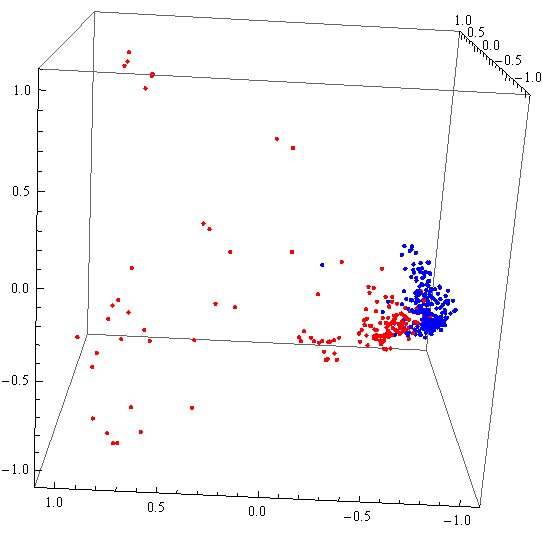}
    \label{fig:humchimp12}
}
\caption{Three-dimensional Molecular Distance Maps  of  150~kbp genomic DNA  sequences from \textit{H.~sapiens} (blue), \textit{P.~troglodytes} (red) using the six distances.}
\label{fig:all_humchimp_3d}
\end{figure}

\begin{table}[H]
\centering
\begin{tabular}{|l|r|r|r|r|c|r|c|}
\hline
& \multicolumn{1}{c|}{$\mathcal D_\alpha$} 
& \multicolumn{1}{c|}{$\mathcal A_\alpha$}
& \multicolumn{1}{c|}{$\mathcal O_\alpha$}
& \multicolumn{1}{c|}{$z$-score sum} 
& \multicolumn{1}{c|}{Rank}
\\
\hline
DSSIM        & $0.167$ & $0.915$ & $0.136$ &   $3.453$   & $1$st  \\ \hline 
Descriptors  & $0.015$ & $0.500$ & $0.101$ &   $-2.593$  & $5$th  \\ \hline 
Euclidean    & $0.037$ & $0.58$  & $0.069$ &   $-2.899$  & $6$th  \\ \hline
Manhattan    & $0.112$ & $0.863$ & $0.108$ &   $1.27$    & $3$rd   \\ \hline 
Pearson      & $0.142$ & $0.714$ & $0.119$ &   $1.339$   & $2$nd   \\ \hline 
Approx. Inf. & $0.075$ & $0.933$ & $0.062$ &   $-0.569$  & $4$th   \\ \hline 
\end{tabular}
\vspace{3mm}
\caption{Summary of quality measures for the performances of six distances (DSSIM, descriptors, Euclidean, Manhattan, Pearson, approximated information distance)  on a dataset of  402 DNA sequences  from \textit{H. sapiens}, chromosome 21 and \textit{P. troglodytes}, chromosome Y.  $\mathcal{D}_\alpha$ is the  correlation to an idealized cluster, $\mathcal{A}_\alpha$ is the silhouette cluster accuracy, and $\mathcal{O}_\alpha$ is the  relative overlap.}
\label{table:humchimp}
\end{table}

The Molecular Distance Maps  in  Figure \ref{fig:all_humchimp_2d} and Figure  \ref{fig:all_humchimp_3d}, of 402 DNA sequences,  suggests  that  several of the distances are able to differentiate even between DNA sequences from closely related organisms.  As seen in Table \ref{table:humchimp},  the Euclidean distance  was again outperformed by other distances, when assessed with the quality measures we described. In this case-study,  we note a change in the distance rankings: DSSIM, which  ranked second previously, now ranks first, while the descriptor distance, which ranked first previously, now ranks  second last. This may be an indication that  descriptor distance, which was designed to  detect pattern differences, may only perform well for analyses of sequences of distantly related organisms while DSSIM, which is sensitive to small differences in  similar images, may be the preferred option for fine-grained analyses at the genus, family and species level.

Further large-scale computational experiments have to be carried out to confirm these preliminary results and establish their validity. Such experiments could provide additional insights regarding the  choice of optimal distance for structural genome comparison in different settings.\\


\begin{backmatter}

\section*{Competing interests}
  The authors declare that they have no competing interests.

\section*{Author's contributions}
RK data collection; data analysis, methodology and result interpretation; manuscript draft; manuscript editing; software design. LK data analysis,  methodology and result interpretation; manuscript draft; manuscript editing. S.Kon data analysis, methodology  and result interpretation; manuscript editing.
S.Kop data analysis, methodology  and result interpretation; manuscript editing. All authors read and approved the final manuscript.

\section*{Acknowledgements}
We thank Yuri Boykov, Lena Gorelick and Olga Veksler for discussions on image descriptors, Stephen Solis for comments on earlier drafts of the manuscript, Genlou Sun for biology expertise, and Nikesh Dattani for some assistance with the NCBI interface.


\bibliographystyle{bmc-mathphys} 
\bibliography{inter-intra}      


\begin{thebibliography}{24}
\ifx \bisbn   \undefined \def \bisbn  #1{ISBN #1}\fi
\ifx \binits  \undefined \def \binits#1{#1}\fi
\ifx \bauthor  \undefined \def \bauthor#1{#1}\fi
\ifx \batitle  \undefined \def \batitle#1{#1}\fi
\ifx \bjtitle  \undefined \def \bjtitle#1{#1}\fi
\ifx \bvolume  \undefined \def \bvolume#1{\textbf{#1}}\fi
\ifx \byear  \undefined \def \byear#1{#1}\fi
\ifx \bissue  \undefined \def \bissue#1{#1}\fi
\ifx \bfpage  \undefined \def \bfpage#1{#1}\fi
\ifx \blpage  \undefined \def \blpage #1{#1}\fi
\ifx \burl  \undefined \def \burl#1{\textsf{#1}}\fi
\ifx \doiurl  \undefined \def \doiurl#1{\textsf{#1}}\fi
\ifx \betal  \undefined \def \betal{\textit{et al.}}\fi
\ifx \binstitute  \undefined \def \binstitute#1{#1}\fi
\ifx \binstitutionaled  \undefined \def \binstitutionaled#1{#1}\fi
\ifx \bctitle  \undefined \def \bctitle#1{#1}\fi
\ifx \beditor  \undefined \def \beditor#1{#1}\fi
\ifx \bpublisher  \undefined \def \bpublisher#1{#1}\fi
\ifx \bbtitle  \undefined \def \bbtitle#1{#1}\fi
\ifx \bedition  \undefined \def \bedition#1{#1}\fi
\ifx \bseriesno  \undefined \def \bseriesno#1{#1}\fi
\ifx \blocation  \undefined \def \blocation#1{#1}\fi
\ifx \bsertitle  \undefined \def \bsertitle#1{#1}\fi
\ifx \bsnm \undefined \def \bsnm#1{#1}\fi
\ifx \bsuffix \undefined \def \bsuffix#1{#1}\fi
\ifx \bparticle \undefined \def \bparticle#1{#1}\fi
\ifx \barticle \undefined \def \barticle#1{#1}\fi
\ifx \bconfdate \undefined \def \bconfdate #1{#1}\fi
\ifx \botherref \undefined \def \botherref #1{#1}\fi
\ifx \url \undefined \def \url#1{\textsf{#1}}\fi
\ifx \bchapter \undefined \def \bchapter#1{#1}\fi
\ifx \bbook \undefined \def \bbook#1{#1}\fi
\ifx \bcomment \undefined \def \bcomment#1{#1}\fi
\ifx \oauthor \undefined \def \oauthor#1{#1}\fi
\ifx \citeauthoryear \undefined \def \citeauthoryear#1{#1}\fi
\ifx \endbibitem  \undefined \def \endbibitem {}\fi
\ifx \bconflocation  \undefined \def \bconflocation#1{#1}\fi
\ifx \arxivurl  \undefined \def \arxivurl#1{\textsf{#1}}\fi
\csname PreBibitemsHook\endcsname

\bibitem{barcoding}
\begin{barticle}
\bauthor{\bsnm{Hebert}, \binits{P.D.}},
\bauthor{\bsnm{Cywinska}, \binits{A.}},
\bauthor{\bsnm{Ball}, \binits{S.L.}}, \betal:
\batitle{Biological identifications through {DNA} barcodes}.
\bjtitle{Proceedings of the Royal Society of London. Series B: Biological
  Sciences}
\bvolume{270}(\bissue{1512}),
\bfpage{313}--\blpage{321}
(\byear{2003})
\end{barticle}
\endbibitem

\bibitem{klee2010biodiversity}
\begin{barticle}
\bauthor{\bsnm{Sirovich}, \binits{L.}},
\bauthor{\bsnm{Stoeckle}, \binits{M.Y.}},
\bauthor{\bsnm{Zhang}, \binits{Y.}}:
\batitle{Structural analysis of biodiversity}.
\bjtitle{PLoS One}
\bvolume{5}(\bissue{2}),
\bfpage{9266}
(\byear{2010})
\end{barticle}
\endbibitem

\bibitem{Jeffrey90}
\begin{barticle}
\bauthor{\bsnm{Jeffrey}, \binits{H.}}:
\batitle{{Chaos Game Representation of gene structure.}}
\bjtitle{Nucleic Acids Research}
\bvolume{18}(\bissue{8}),
\bfpage{2163}--\blpage{2170}
(\byear{1990})
\end{barticle}
\endbibitem

\bibitem{MapOfLife}
\begin{botherref}
\oauthor{\bsnm{Kari}, \binits{L.}},
\oauthor{\bsnm{Hill}, \binits{K.A.}},
\oauthor{\bsnm{Sayem}, \binits{A.S.}},
\oauthor{\bsnm{Karamichalis}, \binits{R.}},
\oauthor{\bsnm{Bryans}, \binits{N.}},
\oauthor{\bsnm{Davis}, \binits{K.}},
\oauthor{\bsnm{Dattani}, \binits{N.S.}}:
Mapping the {S}pace of {G}enomic {S}ignatures.
ArXiv e-prints \url{http://arxiv.org/abs/1406.4105}
({PL}o{S} {O}ne {A}ccepted, {F}eb 10, 2015)
\end{botherref}
\endbibitem

\bibitem{Edwards2002}
\begin{barticle}
\bauthor{\bsnm{Edwards}, \binits{S.}},
\bauthor{\bsnm{Fertil}, \binits{B.}},
\bauthor{\bsnm{Girron}, \binits{A.}},
\bauthor{\bsnm{Deschavanne}, \binits{P.}}:
\batitle{A genomic schism in birds revealed by phylogenetic analysis of {DNA}
  strings}.
\bjtitle{Systematic Biology}
\bvolume{51}(\bissue{4}),
\bfpage{599}--\blpage{613}
(\byear{2002})
\end{barticle}
\endbibitem

\bibitem{HIV_analysis}
\begin{barticle}
\bauthor{\bsnm{Pandit}, \binits{A.}},
\bauthor{\bsnm{Vadlamudi}, \binits{J.}},
\bauthor{\bsnm{Sinha}, \binits{S.}}:
\batitle{Analysis of dinucleotide signatures in {HIV}-1 subtype {B} genomes}.
\bjtitle{Journal of genetics}
\bvolume{92}(\bissue{3}),
\bfpage{403}--\blpage{412}
(\byear{2013})
\end{barticle}
\endbibitem

\bibitem{Deschavanne1999}
\begin{barticle}
\bauthor{\bsnm{Deschavanne}, \binits{P.}},
\bauthor{\bsnm{Giron}, \binits{A.}},
\bauthor{\bsnm{Vilain}, \binits{J.}},
\bauthor{\bsnm{Fagot}, \binits{G.}},
\bauthor{\bsnm{Fertil}, \binits{B.}}:
\batitle{Genomic signature: characterization and classification of species
  assessed by {Chaos Game Representation} of sequences.}
\bjtitle{Molecular Biology and Evolution}
\bvolume{16}(\bissue{10}),
\bfpage{1391}--\blpage{1399}
(\byear{1999})
\end{barticle}
\endbibitem

\bibitem{karlin95}
\begin{barticle}
\bauthor{\bsnm{Gentles}, \binits{A.J.}},
\bauthor{\bsnm{Karlin}, \binits{S.}}:
\batitle{Genome-scale compositional comparisons in eukaryotes}.
\bjtitle{Genome Research}
\bvolume{11}(\bissue{4}),
\bfpage{540}--\blpage{546}
(\byear{2001}).
doi:\doiurl{10.1101/gr.163101}
\end{barticle}
\endbibitem

\bibitem{Deschavanne2000}
\begin{bchapter}
\bauthor{\bsnm{Deschavanne}, \binits{P.}},
\bauthor{\bsnm{Giron}, \binits{A.}},
\bauthor{\bsnm{Vilain}, \binits{J.}},
\bauthor{\bsnm{Dufraigne}, \binits{C.}},
\bauthor{\bsnm{Fertil}, \binits{B.}}:
\bctitle{Genomic signature is preserved in short {DNA} fragments}.
In: \bbtitle{Proceedings of IEEE International Symposium on Bio-Informatics and
  Biomedical Engineering},
pp. \bfpage{161}--\blpage{167}
(\byear{2000})
\end{bchapter}
\endbibitem

\bibitem{ssim}
\begin{barticle}
\bauthor{\bsnm{Wang}, \binits{Z.}},
\bauthor{\bsnm{Bovik}, \binits{A.C.}},
\bauthor{\bsnm{Sheikh}, \binits{H.R.}},
\bauthor{\bsnm{Simoncelli}, \binits{E.P.}}:
\batitle{Image quality assessment: From error visibility to structural
  similarity}.
\bjtitle{IEEE Transactions on Image Processing}
\bvolume{13}(\bissue{4}),
\bfpage{600}--\blpage{612}
(\byear{2004}).
doi:\doiurl{10.1109/TIP.2003.819861}
\end{barticle}
\endbibitem

\bibitem{pearson}
\begin{bbook}
\bauthor{\bsnm{Iversen}, \binits{G.R.}},
\bauthor{\bsnm{Gergen}, \binits{M.}},
\bauthor{\bsnm{Gergen}, \binits{M.M.}}:
\bbtitle{Statistics: The Conceptual Approach}.
\bpublisher{Springer},
\blocation{Berlin Heidelberg}
(\byear{1997})
\end{bbook}
\endbibitem

\bibitem{manhattan}
\begin{bbook}
\bauthor{\bsnm{Krause}, \binits{E.F.}}:
\bbtitle{Taxicab Geometry: An Adventure in non-Euclidean Geometry}.
\bpublisher{Courier Dover Publications},
\blocation{Mineola, New York}
(\byear{2012})
\end{bbook}
\endbibitem

\bibitem{Li2004}
\begin{barticle}
\bauthor{\bsnm{Li}, \binits{M.}},
\bauthor{\bsnm{Chen}, \binits{X.}},
\bauthor{\bsnm{Li}, \binits{X.}},
\bauthor{\bsnm{Ma}, \binits{B.}},
\bauthor{\bsnm{Vitany}, \binits{P.}}:
\batitle{The similarity metric}.
\bjtitle{IEEE Transactions on Information Theory}
\bvolume{50}(\bissue{12}),
\bfpage{3250}--\blpage{3264}
(\byear{2004})
\end{barticle}
\endbibitem

\bibitem{Jeffrey92}
\begin{barticle}
\bauthor{\bsnm{Jeffrey}, \binits{H.}}:
\batitle{Chaos game visualization of sequences}.
\bjtitle{Comput. Graphics}
\bvolume{16}(\bissue{1}),
\bfpage{25}--\blpage{33}
(\byear{1992})
\end{barticle}
\endbibitem

\bibitem{Pride2003}
\begin{barticle}
\bauthor{\bsnm{Pride}, \binits{D.}},
\bauthor{\bsnm{Meinersmann}, \binits{R.}},
\bauthor{\bsnm{Wassenaar}, \binits{T.}},
\bauthor{\bsnm{Blaser}, \binits{M.}}:
\batitle{Evolutionary implications of microbial genome tetranucleotide
  frequency biases}.
\bjtitle{Genome Research}
\bvolume{13}(\bissue{2}),
\bfpage{145}--\blpage{158}
(\byear{2003})
\end{barticle}
\endbibitem

\bibitem{Deschavanne2010}
\begin{barticle}
\bauthor{\bsnm{Deschavanne}, \binits{P.}},
\bauthor{\bsnm{DuBow}, \binits{M.}},
\bauthor{\bsnm{Regeard}, \binits{C.}}:
\batitle{The use of genomic signature distance between bacteriophages and their
  hosts diplays evolutionary relationships and phage growth cycle
  determination}.
\bjtitle{Virology Journal}
\bvolume{7}(\bissue{1}),
\bfpage{163}
(\byear{2010})
\end{barticle}
\endbibitem

\bibitem{Pandit2010}
\begin{barticle}
\bauthor{\bsnm{Pandit}, \binits{A.}},
\bauthor{\bsnm{Sinha}, \binits{S.}}:
\batitle{Using genomic signatures for {HIV}-1 subtyping}.
\bjtitle{BMC Bioinformatics}
\bvolume{11}(\bissue{Suppl 1}),
\bfpage{26}
(\byear{2010})
\end{barticle}
\endbibitem

\bibitem{deza2009}
\begin{bbook}
\bauthor{\bsnm{Deza}, \binits{M.M.}},
\bauthor{\bsnm{Deza}, \binits{E.}}:
\bbtitle{Encyclopedia of Distances}.
\bpublisher{Springer},
\blocation{Berlin Heidelberg}
(\byear{2009})
\end{bbook}
\endbibitem

\bibitem{kruskal}
\begin{barticle}
\bauthor{\bsnm{Kruskal}, \binits{J.}}:
\batitle{Multidimensional scaling by optimizing goodness of fit to a nonmetric
  hypothesis}.
\bjtitle{Psychometrika}
\bvolume{29}(\bissue{1}),
\bfpage{1}--\blpage{27}
(\byear{1964})
\end{barticle}
\endbibitem

\bibitem{gitIntraSupplemental}
\begin{botherref}
Supplemental {M}aterial.
\url{https://github.com/rallis/intraSupplemental_Material}
\end{botherref}
\endbibitem

\bibitem{gitIntraModMap}
\begin{botherref}
\oauthor{\bsnm{Karamichalis}, \binits{R.}}:
Molecular {D}istance {M}ap {I}nteractive {W}ebtool (2014).
\url{https://github.com/rallis/intraMoDMap}
\end{botherref}
\endbibitem

\bibitem{tan2007}
\begin{bchapter}
\bauthor{\bsnm{Pang-Ning}, \binits{T.}},
\bauthor{\bsnm{Steinbach}, \binits{M.}},
\bauthor{\bsnm{Kumar}, \binits{V.}}, \betal:
\bctitle{Introduction to data mining}.
In: \bbtitle{Library of Congress}
(\byear{2006})
\end{bchapter}
\endbibitem

\bibitem{zhao2004empirical}
\begin{barticle}
\bauthor{\bsnm{Zhao}, \binits{Y.}},
\bauthor{\bsnm{Karypis}, \binits{G.}}:
\batitle{Empirical and theoretical comparisons of selected criterion functions
  for document clustering}.
\bjtitle{Machine Learning}
\bvolume{55}(\bissue{3}),
\bfpage{311}--\blpage{331}
(\byear{2004})
\end{barticle}
\endbibitem

\bibitem{silhouette}
\begin{barticle}
\bauthor{\bsnm{Rousseeuw}, \binits{P.J.}}:
\batitle{Silhouettes: A graphical aid to the interpretation and validation of
  cluster analysis}.
\bjtitle{Journal of Computational and Applied Mathematics}
\bvolume{20}(\bissue{0}),
\bfpage{53}--\blpage{65}
(\byear{1987}).
doi:\doiurl{10.1016/0377-0427(87)90125-7}
\end{barticle}
\endbibitem

\end{thebibliography}

\newcommand{\BMCxmlcomment}[1]{}

\BMCxmlcomment{

<refgrp>

<bibl id="B1">
  <title><p>Biological identifications through {DNA} barcodes</p></title>
  <aug>
    <au><snm>Hebert</snm><fnm>PD</fnm></au>
    <au><snm>Cywinska</snm><fnm>A</fnm></au>
    <au><snm>Ball</snm><fnm>SL</fnm></au>
    <au><cnm>others</cnm></au>
  </aug>
  <source>Proceedings of the Royal Society of London. Series B: Biological
  Sciences</source>
  <publisher>The Royal Society</publisher>
  <pubdate>2003</pubdate>
  <volume>270</volume>
  <issue>1512</issue>
  <fpage>313</fpage>
  <lpage>-321</lpage>
</bibl>

<bibl id="B2">
  <title><p>Structural analysis of biodiversity</p></title>
  <aug>
    <au><snm>Sirovich</snm><fnm>L</fnm></au>
    <au><snm>Stoeckle</snm><fnm>MY</fnm></au>
    <au><snm>Zhang</snm><fnm>Y</fnm></au>
  </aug>
  <source>PLoS One</source>
  <publisher>Public Library of Science</publisher>
  <pubdate>2010</pubdate>
  <volume>5</volume>
  <issue>2</issue>
  <fpage>e9266</fpage>
</bibl>

<bibl id="B3">
  <title><p>{Chaos Game Representation of gene structure.}</p></title>
  <aug>
    <au><snm>Jeffrey</snm><fnm>H.</fnm></au>
  </aug>
  <source>Nucleic Acids Research</source>
  <pubdate>1990</pubdate>
  <volume>18</volume>
  <issue>8</issue>
  <fpage>2163</fpage>
  <lpage>-2170</lpage>
  <url>http://view.ncbi.nlm.nih.gov/pubmed/2336393</url>
</bibl>

<bibl id="B4">
  <title><p>Mapping the {S}pace of {G}enomic {S}ignatures.</p></title>
  <aug>
    <au><snm>Kari</snm><fnm>L</fnm></au>
    <au><snm>Hill</snm><fnm>KA</fnm></au>
    <au><snm>Sayem</snm><fnm>AS</fnm></au>
    <au><snm>Karamichalis</snm><fnm>R</fnm></au>
    <au><snm>Bryans</snm><fnm>N</fnm></au>
    <au><snm>Davis</snm><fnm>K</fnm></au>
    <au><snm>Dattani</snm><fnm>NS</fnm></au>
  </aug>
  <source>ArXiv e-prints \url{http://arxiv.org/abs/1406.4105}</source>
  <pubdate>{PL}o{S} {O}ne {A}ccepted, {F}eb 10, 2015</pubdate>
</bibl>

<bibl id="B5">
  <title><p>A genomic schism in birds revealed by phylogenetic analysis of
  {DNA} strings</p></title>
  <aug>
    <au><snm>Edwards</snm><fnm>S.</fnm></au>
    <au><snm>Fertil</snm><fnm>B.</fnm></au>
    <au><snm>Girron</snm><fnm>A.</fnm></au>
    <au><snm>Deschavanne</snm><fnm>P.</fnm></au>
  </aug>
  <source>Systematic Biology</source>
  <pubdate>2002</pubdate>
  <volume>51</volume>
  <issue>4</issue>
  <fpage>599</fpage>
  <lpage>613</lpage>
</bibl>

<bibl id="B6">
  <title><p>Analysis of dinucleotide signatures in {HIV}-1 subtype {B}
  genomes</p></title>
  <aug>
    <au><snm>Pandit</snm><fnm>A</fnm></au>
    <au><snm>Vadlamudi</snm><fnm>J</fnm></au>
    <au><snm>Sinha</snm><fnm>S</fnm></au>
  </aug>
  <source>Journal of genetics</source>
  <publisher>Springer</publisher>
  <pubdate>2013</pubdate>
  <volume>92</volume>
  <issue>3</issue>
  <fpage>403</fpage>
  <lpage>-412</lpage>
</bibl>

<bibl id="B7">
  <title><p>Genomic signature: characterization and classification of species
  assessed by {Chaos Game Representation} of sequences.</p></title>
  <aug>
    <au><snm>Deschavanne</snm><fnm>P.</fnm></au>
    <au><snm>Giron</snm><fnm>A.</fnm></au>
    <au><snm>Vilain</snm><fnm>J.</fnm></au>
    <au><snm>Fagot</snm><fnm>G.</fnm></au>
    <au><snm>Fertil</snm><fnm>B.</fnm></au>
  </aug>
  <source>Molecular Biology and Evolution</source>
  <pubdate>1999</pubdate>
  <volume>16</volume>
  <issue>10</issue>
  <fpage>1391</fpage>
  <lpage>-1399</lpage>
</bibl>

<bibl id="B8">
  <title><p>Genome-Scale Compositional Comparisons in Eukaryotes</p></title>
  <aug>
    <au><snm>Gentles</snm><fnm>AJ</fnm></au>
    <au><snm>Karlin</snm><fnm>S</fnm></au>
  </aug>
  <source>Genome Research</source>
  <pubdate>2001</pubdate>
  <volume>11</volume>
  <issue>4</issue>
  <fpage>540</fpage>
  <lpage>546</lpage>
</bibl>

<bibl id="B9">
  <title><p>Genomic signature is preserved in short {DNA} fragments</p></title>
  <aug>
    <au><snm>Deschavanne</snm><fnm>P</fnm></au>
    <au><snm>Giron</snm><fnm>A</fnm></au>
    <au><snm>Vilain</snm><fnm>J</fnm></au>
    <au><snm>Dufraigne</snm><fnm>C</fnm></au>
    <au><snm>Fertil</snm><fnm>B</fnm></au>
  </aug>
  <source>Proceedings of IEEE International Symposium on Bio-Informatics and
  Biomedical Engineering</source>
  <pubdate>2000</pubdate>
  <fpage>161</fpage>
  <lpage>-167</lpage>
</bibl>

<bibl id="B10">
  <title><p>Image quality assessment: From error visibility to structural
  similarity</p></title>
  <aug>
    <au><snm>Wang</snm><fnm>Z</fnm></au>
    <au><snm>Bovik</snm><fnm>A.C.</fnm></au>
    <au><snm>Sheikh</snm><fnm>H.R.</fnm></au>
    <au><snm>Simoncelli</snm><fnm>E.P.</fnm></au>
  </aug>
  <source>IEEE Transactions on Image Processing</source>
  <pubdate>2004</pubdate>
  <volume>13</volume>
  <issue>4</issue>
  <fpage>600</fpage>
  <lpage>612</lpage>
</bibl>

<bibl id="B11">
  <title><p>Statistics: The conceptual approach</p></title>
  <aug>
    <au><snm>Iversen</snm><fnm>GR</fnm></au>
    <au><snm>Gergen</snm><fnm>M</fnm></au>
    <au><snm>Gergen</snm><fnm>MM</fnm></au>
  </aug>
  <publisher>Berlin Heidelberg: Springer</publisher>
  <pubdate>1997</pubdate>
</bibl>

<bibl id="B12">
  <title><p>Taxicab Geometry: An adventure in non-Euclidean
  geometry</p></title>
  <aug>
    <au><snm>Krause</snm><fnm>EF</fnm></au>
  </aug>
  <publisher>Mineola, New York: Courier Dover Publications</publisher>
  <pubdate>2012</pubdate>
</bibl>

<bibl id="B13">
  <title><p>The similarity metric</p></title>
  <aug>
    <au><snm>Li</snm><fnm>M.</fnm></au>
    <au><snm>Chen</snm><fnm>X.</fnm></au>
    <au><snm>Li</snm><fnm>X.</fnm></au>
    <au><snm>Ma</snm><fnm>B.</fnm></au>
    <au><snm>Vitany</snm><fnm>P.</fnm></au>
  </aug>
  <source>IEEE Transactions on Information Theory</source>
  <pubdate>2004</pubdate>
  <volume>50</volume>
  <issue>12</issue>
  <fpage>3250</fpage>
  <lpage>3264</lpage>
</bibl>

<bibl id="B14">
  <title><p>Chaos game visualization of sequences</p></title>
  <aug>
    <au><snm>Jeffrey</snm><fnm>H.</fnm></au>
  </aug>
  <source>Comput. Graphics</source>
  <pubdate>1992</pubdate>
  <volume>16</volume>
  <issue>1</issue>
  <fpage>25</fpage>
  <lpage>33</lpage>
</bibl>

<bibl id="B15">
  <title><p>Evolutionary implications of microbial genome tetranucleotide
  frequency biases</p></title>
  <aug>
    <au><snm>Pride</snm><fnm>D.</fnm></au>
    <au><snm>Meinersmann</snm><fnm>R.</fnm></au>
    <au><snm>Wassenaar</snm><fnm>T.</fnm></au>
    <au><snm>Blaser</snm><fnm>M.</fnm></au>
  </aug>
  <source>Genome Research</source>
  <pubdate>2003</pubdate>
  <volume>13</volume>
  <issue>2</issue>
  <fpage>145</fpage>
  <lpage>158</lpage>
</bibl>

<bibl id="B16">
  <title><p>The use of genomic signature distance between bacteriophages and
  their hosts diplays evolutionary relationships and phage growth cycle
  determination</p></title>
  <aug>
    <au><snm>Deschavanne</snm><fnm>P.</fnm></au>
    <au><snm>DuBow</snm><fnm>M.</fnm></au>
    <au><snm>Regeard</snm><fnm>C.</fnm></au>
  </aug>
  <source>Virology Journal</source>
  <publisher>BioMed Central Ltd</publisher>
  <pubdate>2010</pubdate>
  <volume>7</volume>
  <issue>1</issue>
  <fpage>163</fpage>
</bibl>

<bibl id="B17">
  <title><p>Using genomic signatures for {HIV}-1 subtyping</p></title>
  <aug>
    <au><snm>Pandit</snm><fnm>A.</fnm></au>
    <au><snm>Sinha</snm><fnm>S.</fnm></au>
  </aug>
  <source>BMC Bioinformatics</source>
  <publisher>BioMed Central Ltd</publisher>
  <pubdate>2010</pubdate>
  <volume>11</volume>
  <issue>Suppl 1</issue>
  <fpage>S26</fpage>
</bibl>

<bibl id="B18">
  <title><p>Encyclopedia of distances</p></title>
  <aug>
    <au><snm>Deza</snm><fnm>MM</fnm></au>
    <au><snm>Deza</snm><fnm>E</fnm></au>
  </aug>
  <publisher>Berlin Heidelberg: Springer</publisher>
  <pubdate>2009</pubdate>
</bibl>

<bibl id="B19">
  <title><p>Multidimensional scaling by optimizing goodness of fit to a
  nonmetric hypothesis</p></title>
  <aug>
    <au><snm>Kruskal</snm><fnm>J.</fnm></au>
  </aug>
  <source>Psychometrika</source>
  <pubdate>1964</pubdate>
  <volume>29</volume>
  <issue>1</issue>
  <fpage>1</fpage>
  <lpage>27</lpage>
  <url>http://ideas.repec.org/a/spr/psycho/v29y1964i1p1-27.html</url>
</bibl>

<bibl id="B20">
  <title><p>Supplemental {M}aterial</p></title>
  <source>\url{https://github.com/rallis/intraSupplemental_Material}</source>
</bibl>

<bibl id="B21">
  <title><p>Molecular {D}istance {M}ap {I}nteractive {W}ebtool
  (2014)</p></title>
  <aug>
    <au><snm>Karamichalis</snm><fnm>R.</fnm></au>
  </aug>
  <source>\url{https://github.com/rallis/intraMoDMap}</source>
</bibl>

<bibl id="B22">
  <title><p>Introduction to data mining</p></title>
  <aug>
    <au><snm>Pang Ning</snm><fnm>T</fnm></au>
    <au><snm>Steinbach</snm><fnm>M</fnm></au>
    <au><snm>Kumar</snm><fnm>V</fnm></au>
    <au><cnm>others</cnm></au>
  </aug>
  <source>Library of Congress</source>
  <pubdate>2006</pubdate>
</bibl>

<bibl id="B23">
  <title><p>Empirical and theoretical comparisons of selected criterion
  functions for document clustering</p></title>
  <aug>
    <au><snm>Zhao</snm><fnm>Y</fnm></au>
    <au><snm>Karypis</snm><fnm>G</fnm></au>
  </aug>
  <source>Machine Learning</source>
  <publisher>Springer</publisher>
  <pubdate>2004</pubdate>
  <volume>55</volume>
  <issue>3</issue>
  <fpage>311</fpage>
  <lpage>-331</lpage>
</bibl>

<bibl id="B24">
  <title><p>Silhouettes: A graphical aid to the interpretation and validation
  of cluster analysis</p></title>
  <aug>
    <au><snm>Rousseeuw</snm><fnm>PJ</fnm></au>
  </aug>
  <source>Journal of Computational and Applied Mathematics</source>
  <pubdate>1987</pubdate>
  <volume>20</volume>
  <issue>0</issue>
  <fpage>53</fpage>
  <lpage>65</lpage>
  <url>http://www.sciencedirect.com/science/article/pii/0377042787901257</url>
</bibl>

</refgrp>
} 

\end{backmatter}

\end{document}